\documentclass[a4paper,11pt]{article}
\pdfoutput=1 % if your are submitting a pdflatex (i.e. if you have
\usepackage{jcappub} % for details on the use of the package, please
\usepackage[T1]{fontenc} % if needed
\usepackage[utf8]{inputenc}
\usepackage{xcolor}
\usepackage{graphicx}
\usepackage{bm}
\usepackage{ulem}
\usepackage{multirow, array, float, tabularx}
\usepackage{amsmath}
\PassOptionsToPackage{hyphens}{url}

\newcommand{\LCDM}{$\Lambda$CDM}
\newcommand{\hinvmpc}{\,h^{-1}\rm Mpc}

\newcommand{\lcdm}{$\Lambda{\rm CDM}$}
\newcommand{\wcdm}{$w {\rm CDM}$}

\newcommand{\omgeom}{\Omega_M^{\rm geom}}
\newcommand{\omgrow}{\Omega_M^{\rm grow}}
\newcommand{\dom}{\Delta\Omega_M}
\newcommand{\dw}{\Delta w}
\newcommand{\wgeom}{w^{\rm geom}}
\newcommand{\wgrow}{w^{\rm grow}}
\renewcommand{\arraystretch}{1.25}

\newcommand{\ds}{\displaystyle}

\graphicspath{{./figs/}}

\title{\boldmath A Test of the Standard Cosmological Model with Geometry and Growth}

\author[a,b,1]{Uendert Andrade\note{Corresponding author.}}
\author[b,c,d]{, Dhayaa Anbajagane}
\author[a]{, Rodrigo von Marttens}
\author[b,c,e]{, Dragan Huterer}
\author[a]{and Jailson Alcaniz}

\affiliation[a]{Observat\'orio Nacional, Rio de Janeiro, RJ 20921-400, Brazil}
\affiliation[b]{Department of Physics, University of Michigan, 450 Church St, Ann Arbor, MI 48109-1040}
\affiliation[c]{Leinweber Center for Theoretical Physics, University of Michigan, 450 Church St, Ann Arbor, MI 48109-1040}
\affiliation[d]{Department of Astronomy and Astrophysics, University of Chicago, 5640 S. Ellis Ave, Chicago, IL 60637}
\affiliation[e]{Max-Planck-Institut f\"ur Astrophysik, Karl-Schwarzschild-Str.\ 1, 85748 Garching, Germany}

\emailAdd{uendertandrade@on.br}
\emailAdd{dhayaa@umich.edu}
\emailAdd{rodrigovonmarttens@gmail.com}
\emailAdd{huterer@umich.edu}
\emailAdd{alcaniz@on.br}

\abstract{We perform a general test of the \lcdm\ and \wcdm\ cosmological models by comparing constraints on the geometry of the expansion history to those on the growth of structure. Specifically, we split the total matter energy density, $\Omega_M$, and (for \wcdm) dark energy equation of state, $w$, into two parameters each: one that captures the geometry, and another that captures the growth. We constrain our split models using current cosmological data, including type Ia supernovae, baryon acoustic oscillations, redshift space distortions, gravitational lensing, and cosmic microwave background (CMB) anisotropies. We focus on two tasks: (i) constraining deviations from the standard model, captured by the parameters $\dom \equiv \omgrow-\omgeom$ and $\dw \equiv \wgrow-\wgeom$, and (ii) investigating whether the $S_8$ tension between the CMB and weak lensing can be translated into a tension between geometry and growth, i.e. $\dom \neq 0$, $\dw \neq 0$. In both the split \lcdm\ and \wcdm\ cases, our results from combining all data are consistent with $\dom = 0$ and $\dw = 0$.
If we omit BAO/RSD data and constrain the split \wcdm\ cosmology, we find the data prefers $\dw<0$ at $3.6\sigma$ significance and $\dom>0$ at  $4.2\sigma$ evidence.
We also find that for both CMB and weak lensing, $\dom$ and $S_8$ are correlated, with CMB showing a slightly stronger correlation. The general broadening of the contours in our extended model does alleviate the $S_8$ tension, but the allowed nonzero values of $\dom$ do not encompass the $S_8$ values that would point toward a mismatch between geometry and growth as the origin of the tension.}

\begin{document}
\maketitle
\flushbottom

\section{Introduction}
\label{sec:intro}

The standard $\Lambda$ cold dark matter ($\Lambda$CDM) cosmological model has been spectacularly successful in fitting modern observations \citep{Frieman:2008sn,Weinberg:2012es,Huterer:2017buf}. Nevertheless, the search for \textit{departures} from the \LCDM\ model is a frontier in cosmology, as such a finding would shed significant light on the physics behind cosmic acceleration. 
Recent hints for such tensions, such as the $\sim$5-sigma evidence for the difference in the Hubble constant $H_0$ measured by the local distance ladder and that measured by high-redshift probes~\cite{Aghanim:2018eyx,Riess:2020sih}, as well as the weaker but still interesting tension between the amplitude of mass fluctuations $\sigma_8$ measured by weak gravitational lensing \citep{Abbott:2017wau, Asgari:2020wuj} and the cosmic microwave background \citep{Aghanim:2018eyx,Aiola:2020azj}, have therefore elicited much interest in the field \cite{DiValentino:2020vvd}.

Assuming General Relativity (GR) and a smooth dark energy without anisotropic stresses (but with arbitrary and possibly time-dependent equation of state), the expansion history fully determines the growth of cosmic structure.  In particular, each Fourier mode $k$ of the growth of linear density fluctuations $\delta\equiv\delta\rho_M/\rho_M$ on sub-horizon scales evolves via
\begin{equation}
\ddot\delta + 2H\dot\delta-4\pi G\rho_M \delta = 0,
\label{eq:growth}
\end{equation}
where $H$ is the Hubble parameter and dots are derivatives with respect to time. The expansion history $H(t)$, along with the matter density $\rho_M(t)$, therefore completely determine the evolution of the perturbations, $\delta(t)$. In modified gravity, however, Eq.~(\ref{eq:growth}) may no longer hold, and the growth of structure may be governed by a \textit{different} equation that is model dependent and that may, for example, couple different modes even on linear scales. 

Therefore, comparing measurements of geometric quantities with those describing the growth of structure is a particularly insightful stress-test of the  standard cosmology \cite{Ishak:2005zs,Linder:2005in,Knox:2005rg,Bertschinger:2006aw,Huterer:2006mva}. 
For example, given constraints on the initial conditions (power spectrum shape and amplitude), very precise distance measurements from e.g.\ type Ia supernovae (SN Ia) and baryon acoustic oscillations (BAO) predict the convergence power spectrum measured by weak lensing probes. In this scenario, the weak lensing signal depends on the late-time growth of structure which in turn is precisely determined by distance measurements. An equivalent statement is  that  the simple structure of smooth dark-energy models allows their falsifiability, and allows for tight predictions about quantities that can be measured with current or future surveys \cite{Mortonson:2008qy,Mortonson:2009hk,Vanderveld:2012ec,Miranda:2017mnw,Raveri:2019mxg}. While there exist a number of ways to test for new physics, separately constraining the geometry and growth aspects of the cosmological theory seems particularly promising because a key signature of many modified gravity theories is a mismatch between geometry and growth.

The goal of this work is to apply the geometry-growth split to the latest cosmological data. We apply the split to flat cosmological models with either the cosmological constant (\LCDM) or dark energy of an arbitrary but constant equation of state (\wcdm). At the parameter level, we double the set of late-universe parameter(s) that describe dark energy in order to have one set that determines geometry and another that determines growth of structure. This method, proposed and applied to early data by \cite{Wang:2007fsa}, and then extended by \cite{Ruiz:2014hma, Bernal:2015zom} and, most recently, \cite{Muir:2020puy} and \cite{Ruiz-Zapatero:2021rzl}, is particularly effective because of the aforementioned expectation that beyond-\wcdm\ and modified-gravity signatures would appear as a departure from the expected agreement between geometry and growth observables. Moreover, the parameter-split method  makes no assumptions about the nature of beyond-\LCDM\ physics (beyond the way in which the geometry-growth mismatch is parameterized), and is therefore fairly model-independent. Finally, a separate parameterization of the geometry and growth theory components is particularly useful for probes, such as the cosmic microwave background (CMB), that contain information on both geometry and growth and thus cannot be placed into just one category. We also caution, however, that the detection of any discrepancies between geometry and growth parameters would not automatically imply a departure from GR, as the systematics of each individual probe could affect our results as well.

This paper is organized as follows. In Sec.~\ref{sec:split_def} we define our parameter split for each probe considered in this work. In Sec.~\ref{sec:results}, we describe our results for the marginalized posteriors of all relevant parameters, in both the \LCDM\ and \wcdm\ cases, and detail whether the geometry-growth split can resolve the $S_8$ tension. We discuss our results in Sec.~\ref{sec:discussion}, and summarize our findings in Sec.~\ref{sec:concl}.

\section{Geometry and Growth split}\label{sec:split_def}

Our goal is to constrain the late-universe cosmological parameter sector corresponding to dark energy, while separately using information from either geometry or growth. To do this, we ``split'' the dark-energy parameters by defining a separate one each of geometry and growth. The geometry sector is probed by measurements of distances and volumes, while the growth sector is probed by the evolution of perturbations.

To enable the geometry-growth split in the flat \lcdm\ cosmological model, we take the matter density relative to critical --- which normally one parameter, $\Omega_M$ --- and duplicate it into two parameters: $\omgeom$ and $\omgrow$. In a flat model, the energy density of dark energy is given by $\Omega_\Lambda=1-\Omega_M$, so our split on $\Omega_M$ is a split on $\Omega_\Lambda$. In the analogous scenario the flat \wcdm\ model, the equation-of-state of dark energy, $w$, is also described by two parameters, $\wgeom$ and $\wgrow$. Therefore, our cosmological analysis is specified by
\begin{flalign}
  \qquad \bullet\ \mbox{split\,\,} \Lambda{\rm CDM:} &\quad \{\omgeom, \omgrow, \{p_i\}\}&  \label{paramsom} \\[0.2cm]
  \qquad \bullet\ \mbox{split\,\,} w{\rm CDM:} &\quad \{\omgeom, \omgrow, \wgeom, \wgrow, \{p_i\}\} \label{paramsomw}
\end{flalign}
where $\{p_i\}$ are other, standard cosmological parameters described below in Sec.~\ref{sec:results}.

The remaining task is the modelling of theoretical quantities with the split parameterization in the \lcdm\ and \wcdm\ models. The choice of model is unambiguous for some quantities, e.g.\ the SN Ia distances contain only information about geometry\footnote{Ignoring the very small effect of weak gravitational lensing on SN Ia.}. However, the geometry-growth split can be highly ambiguous for other quantities, such as the baryon acoustic oscillations (BAO), or the various observational aspects of the cosmic microwave background (CMB) anisotropies \citep{Wang:2007fsa}. In the following subsections, we describe the cosmological probes that we use, and how we split their theoretical description into geometrical and growth pieces.

\subsection{CMB split}\label{sec:CMB}

\begin{figure*}
    \centering
    \includegraphics[width=0.49\textwidth, trim={0.6cm 1cm 0.4cm 0.7cm}, clip]{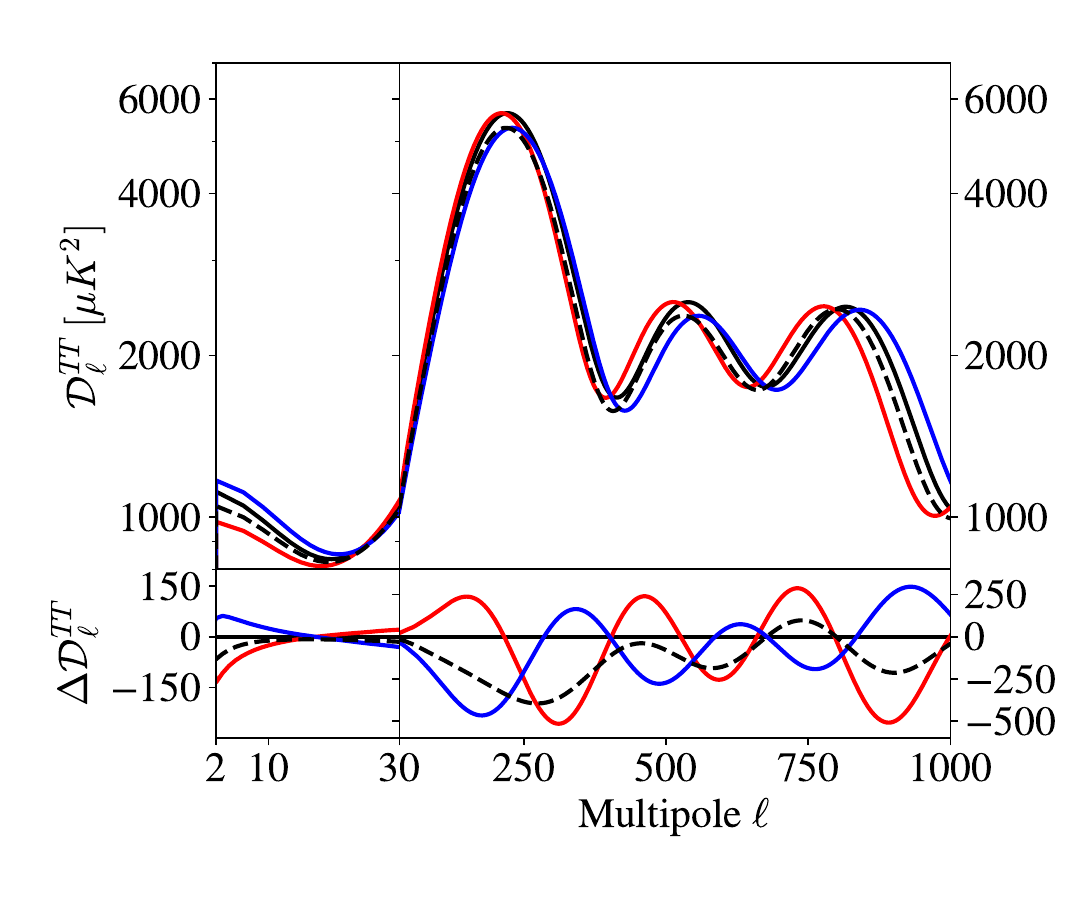}
    \includegraphics[width=0.49\textwidth, trim={0.6cm 1cm 0.4cm 0.7cm}, clip]{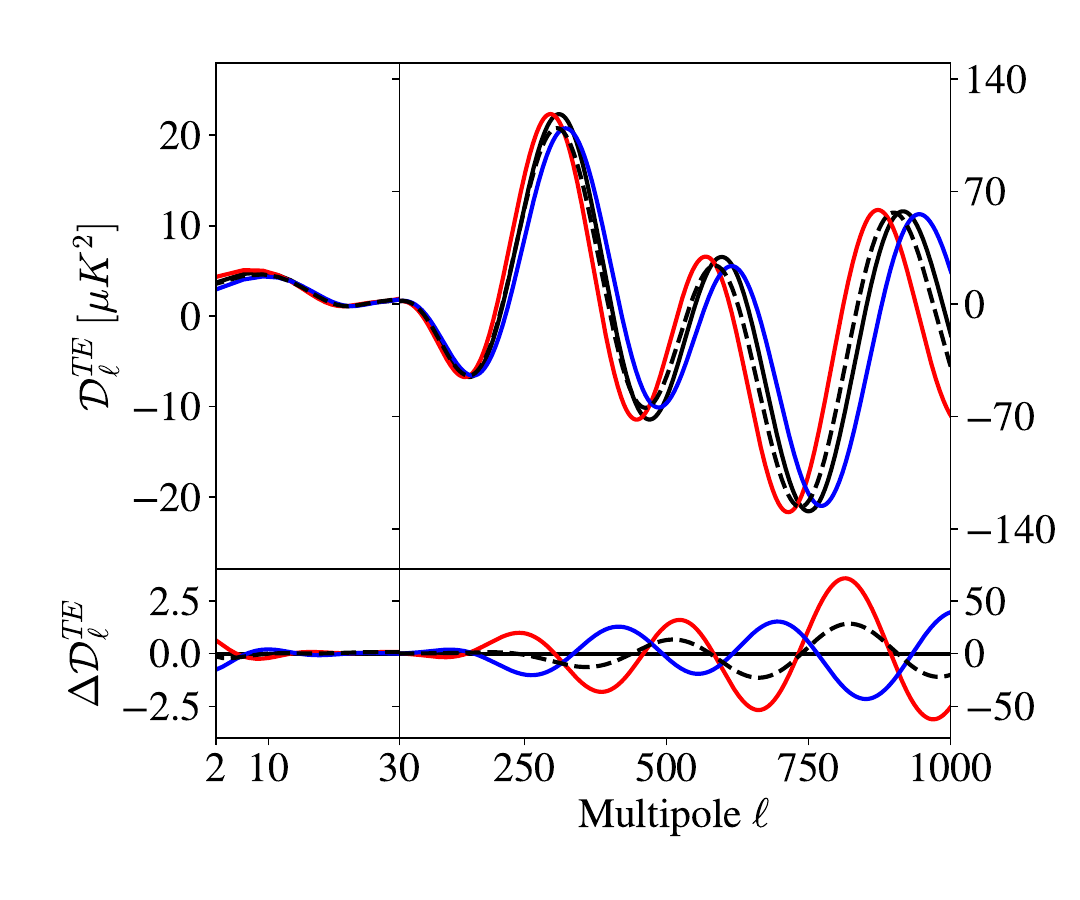}
    \includegraphics[width=0.49\textwidth, trim={0.6cm 1cm 0.4cm 0.7cm}, clip]{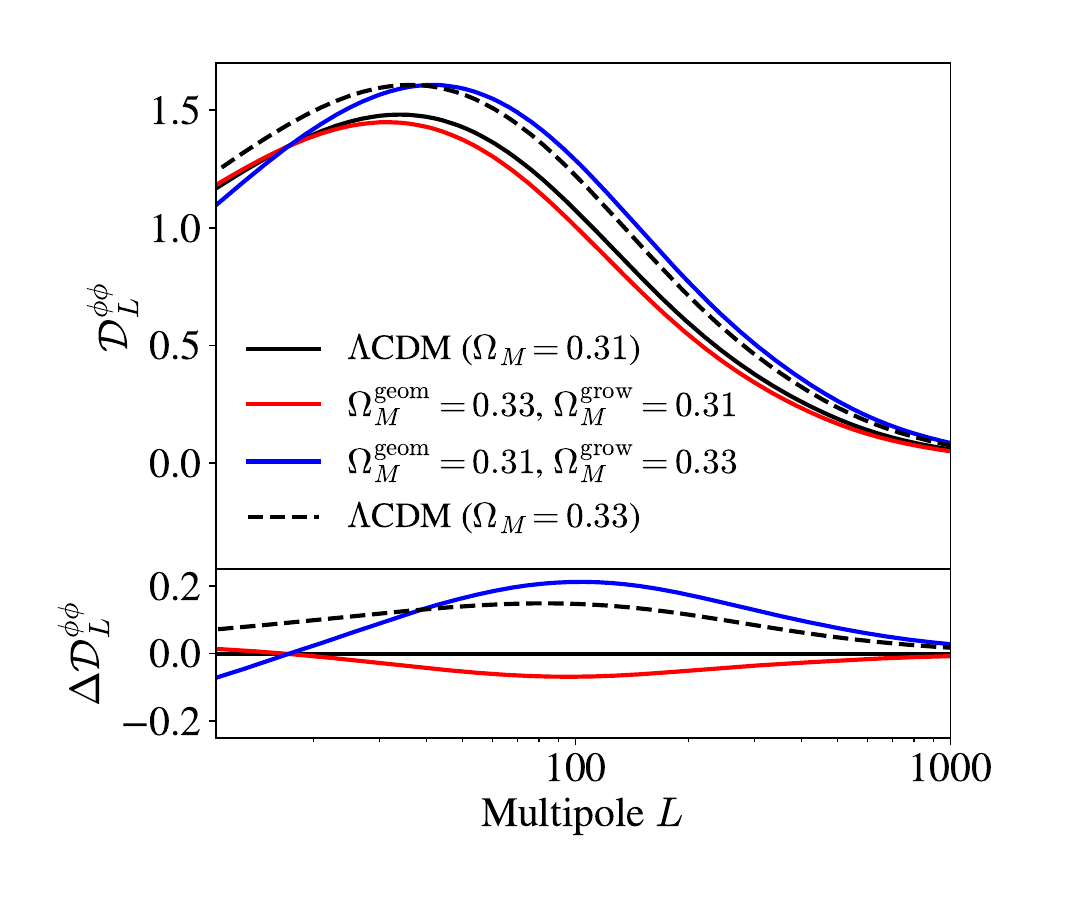}
    \includegraphics[width=0.49\textwidth, trim={0.6cm 1cm 0.4cm 0.7cm}, clip]{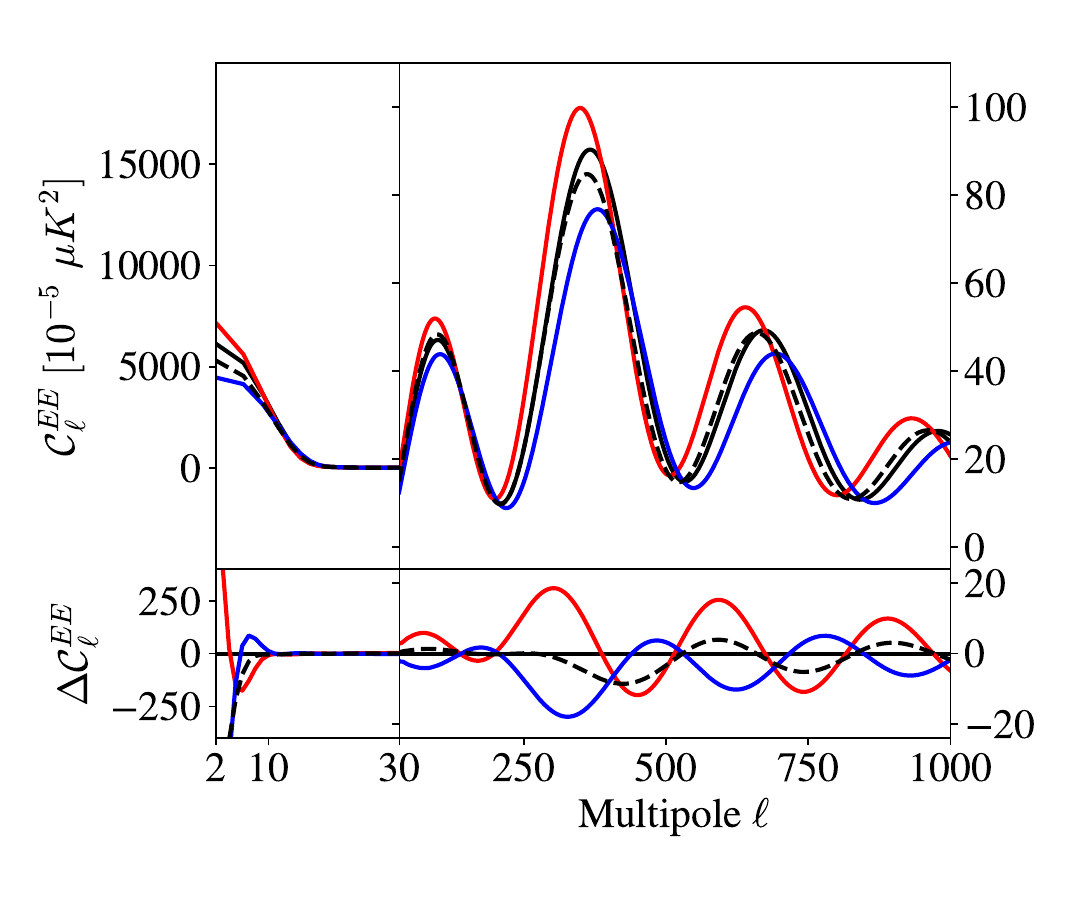}
    \caption{Variations in the CMB anisotropy power spectra (at $z = 0$) due to changes of $\omgeom$ and $\omgrow$. Going clockwise from the top left plot, the plots show the temperature auto-correlation (TT), the temperature-polarization cross-correlation (TE), the polarization auto-correlation (EE), and the Lensing auto-correlation power spectra ($\phi\phi$). Three of the four panels feature the x-axis split in two ranges, with the range left of the vertical line corresponding to the y-axis scale on the left, and one to the right of the vertical line corresponding to the y-axis scale on the right. In all panels,
    the black solid curve corresponds to our fiducial model with $\omgeom=\omgrow=0.31$ while the red curve corresponds to the deviations in the geometry component ($\Omega_c^{geom}=0.33$ and $\omgrow=0.31$), and the blue curve corresponds to deviations in the growth components ($\omgeom=0.31$ and $\omgrow=0.33$). The black dotted curve corresponds to increasing the parameters by the same amount: $\omgeom=\omgrow=0.33$. Note that the red and blue curve are largely out of phase, indicating the CMB's strong sensitivity to the \textit{difference} between $\omgeom$ and $\omgrow$. }
    \label{fig:cmb_tt_split}
\end{figure*}

The temperature anisotropy power spectrum is given by

\begin{equation}
    C_{\ell}^{TT} = \frac{1}{2\pi^2} \int \frac{dk}{k} \, \Theta^2(k, z=0) \, \mathcal{P}_R(k).
\end{equation}
Here $\mathcal{P}_R(k) = A_s (k/k_{0})^{(n_s -1)}$ is the primordial curvature power spectrum, with amplitude $A_S$ and scalar spectral index $n_s$. Moreover $\Theta^2(k,z=0)$ is the transfer function,

\begin{equation}
    \Theta(k,z=0) = \int dz' S_T(k, z') j_{\ell}[k \chi(z')] ,
    \label{eq:transfer_fun}
\end{equation}
where $j_{\ell}$ is spherical Bessel function, $S_T$ is the source function obtained by solving the coupled Einstein-Boltzmann equations, and $\chi$ is the radial distance.

We classify the source function $S_T$ as a growth quantity, and therefore use the parameters $\omgrow$ and $\wgrow$ in its calculation. The Bessel function in Eq.~(\ref{eq:transfer_fun}) describes the geometrical projection, so we model the radial distance $\chi$ here with geometrical parameters, $\omgeom$ and $\wgeom$. An analogous choice for the split is also made for the EE, TE and lensing spectra.  

Finally, we note that throughout this paper we assume that the baryon density, $\Omega_B$, is the same for the geometry and growth sector, and therefore split only the cold-dark-matter density relative to critical density, $\Omega_c = \Omega_M - \Omega_B$, into the geometry and growth sectors\footnote{Whether a split in $\Omega_M$ is implemented as a split in $\Omega_c$ (as in this paper), or that in $\Omega_B$ \cite{Chu:2004qx}, or both, represents a choice, one of many when implementing the geometry-growth test. However note that this particular choice is not too important as we are mainly interested in the geometry-growth mismatch in the late-universe physics, corresponding to the dark energy density $1-\Omega_M= 1-\Omega_c-\Omega_B$.}.

While Eq.~(\ref{eq:transfer_fun}) is written in terms of redshift, the publicly available Boltzmann codes solve pertubative equations in terms of the conformal time $\eta$. Both the geometrical and the growth quantities are calculated using the conformal time: geometry (e.g.\ distances) depends on the scale factor which in terms depends on time, while the growth rate  also requires time; see  Eq.~\eqref{eq:growth}. Therefore, to enable a geometry-growth split, it is necessary to split the conformal time into two separately evolving quantities, one controlling the geometry, $\eta^{\rm geom}$, and the other, $\eta^{\rm grow}$, controlling the growth. In practice this means that, at a fixed redshift, the two conformal times, $\eta^{\rm geom}$ and $\eta^{\rm grow}$, will generally be different \citep{Wang:2007fsa}. 

Figure \ref{fig:cmb_tt_split} illustrates how the CMB observables respond to the split parameterization. The top left panel shows the CMB temperature angular power spectrum when we split $\Omega_M$ into  $\{\omgrow, \omgeom\}$. The black solid curve is our fiducial model with $\omgrow=\omgeom=0.31$. The red curve shows how the TT spectrum changes as we increase the geometric matter density to $ \omgeom=0.33$ (holding $ \omgrow=0.31$ unchanged). We observe a horizontal displacement because, with the higher matter density, the distance to the last scattering surface decreases, shifting the angular power spectrum to larger angular scales (lower multipoles). The blue curve shows the case when we change the growth matter parameter to $ \omgrow=0.33$ (holding $ \omgeom=0.31$ unchanged). Now the angular power spectrum is shifted to smaller angular scales because the radiation-matter transition happens earlier, implying a smaller sound horizon and thus a smaller subtended angle.  Finally, when \textit{both} $\omgrow$ and $\omgeom$ are increased to $0.33$, the two aforementioned effects partly cancel, as the black doted curve shows.

The top right and the bottom right panel of Fig.~\ref{fig:cmb_tt_split} show that the effects of the split on the TE and EE polarization is similar to that of the TT result described above. Finally, the bottom left panel shows the CMB lensing power spectrum. Lensing distinguishes itself from the TT, TE, and EE spectra in that it is much more sensitive to the growth than the geometry parameters, and this can be seen in how lensing amplitude increases with $\omgrow$.

All panels in Fig.~\ref{fig:cmb_tt_split} indicate that relative changes in the various CMB power spectra when $\omgeom$ is increased are mutually out of phase to those when $\omgrow$ is increased. Therefore, we expect that the \textit{difference} between the two parameters,  $\Delta \Omega_M \equiv \omgrow - \omgeom$, will be tightly constrained. Our results below will confirm this.

For the CMB part of our analysis, we employ the most recent 2018 Planck data\footnote{\url{http://pla.esac.esa.int/pla/\#cosmology}}, which rely on the temperature, polarization, and lensing maps. More precisely, we use the following likelihood codes: \textsc{Commander} for TT spectrum with $2\leq\ell<30$, \textsc{SimAll} for EE spectrum with $2\leq\ell<30$ and finally \textsc{PlickTT,TE,EE} for TT spectrum with $30\leq\ell<2500$ as well as TE and EE spectra with $30\leq\ell\lesssim 2000$. For lensing, we use the standard Planck likelihood obtained from the power spectrum reconstruction with the conservative multipole\footnote{We follow the usual convention of using $L$ for lensing multipoles instead of $\ell$.} range $8\leq L\leq400$. Further details about the likelihoods for the Planck TTTEEE and lensing power spectra can be found in Ref.~\cite{Aghanim:2019ame} and Ref.~\cite{Aghanim:2018oex} respectively.

\subsection{Weak Lensing split}\label{sec:WL}

Weak gravitational lensing is another powerful probe of both geometry and growth. Here the information is usually compressed into the real-space two-point correlation function, $\xi_{\pm}^{ij}$, where $i$ and $j$ refer to tomographic redshift bins. The theoretical expectation for the weak lensing power spectrum is
\begin{equation}
    \xi_{\pm}^{ij} = \frac{1}{2\pi} \int d\ell \ell P_k^{ij}(\ell) J_{0,4}(\ell \theta) \,,
\end{equation} 
where $J_{0}(J_{4})$ is the zeroth (fourth)-order Bessel function used in the expression for $\xi_{+}^{ij}(\xi_{-}^{ij})$, and $P_\kappa^{ij}(\ell) $ is the convergence power spectrum which is in turn related to the matter power spectrum, $P_{\delta}$. In the Limber approximation, the convergence power spectrum is given by, 
\begin{equation}\label{eq:convergence}
    P_\kappa^{ij}(\ell) = \int_0^{\chi_H} d\chi \, \frac{q_i(\chi) \, q_j(\chi)}{\chi^2} P_{\delta}\left(\frac{\ell}{\chi(z)}, z \right)\, ,
\end{equation}
where $q_i(\chi)$ is the lensing efficiency function,
\begin{equation}
    q_i(\chi) = \frac{3H_0^2\Omega_m}{2c^2} \frac{\chi}{a(\chi)} \int_{\chi}^{\chi_H} d\chi' n_i(\chi')\, \frac{\chi'-\chi}{\chi'}\,,
\end{equation}
and $ \chi_H$ is the horizon distance. The parameter $n_i(\chi)$ is the distribution of galaxies in each redshift bin, normalized to $\int_0^{\chi_H} n_i(\chi)d\chi = 1 $, while $a(\chi)$ is the scale factor. 

Rewriting the convergence power spectrum in terms of redshift, we have

\begin{align} \label{eq:convergence_z}
P_\kappa^{ij}(\ell) & = \frac{9}{4}\Omega^2_m H_0^4 \int^{\infty}_0 dz
(1+z)^2 \left[\frac{d\chi(z)}{dz}\right] g_i(z) g_j(z)
P_{\delta}\left(\frac{\ell}{\chi(z)}, z \right),\\
g_i(z) &\equiv \int^{\infty}_z dz'n_i(z')
\left[\frac{\chi(z')-\chi(z)}{\chi(z')}\right].\label{eq:gz}
\end{align}

The quantity $g_i(z)$ determines the geometry for the bending of light, so we choose all quantities in Eq.~(\ref{eq:gz}) to be described by geometry parameters. Since the radial distance $\chi$ is a geometrical quantity, so is the wavenumber that enters the matter power spectrum, $k=\ell/\chi$. This makes the $k$-dependent terms in the power spectrum --- the power law, and the redshift-zero transfer function --- geometry as well.   However the linear-growth term that enters the matter power spectrum is classified as growth 
as it comes from the solving the growth equation. Additionally, the prefactor $\Omega_M^2$ outside of the integral in Eq.~(\ref{eq:convergence_z}) is treated as growth \cite{Wang:2007fsa,Matilla:2017rmu}. Finally, to obtain the non-linear matter power spectrum we use the {\sc HMCode} recipe, and consider its input to be entirely growth \cite{Muir:2020puy}.

\begin{figure}
    \centering
    \includegraphics[width=0.48\textwidth, trim={0.7cm 1cm 1.7cm 0.7cm}, clip]{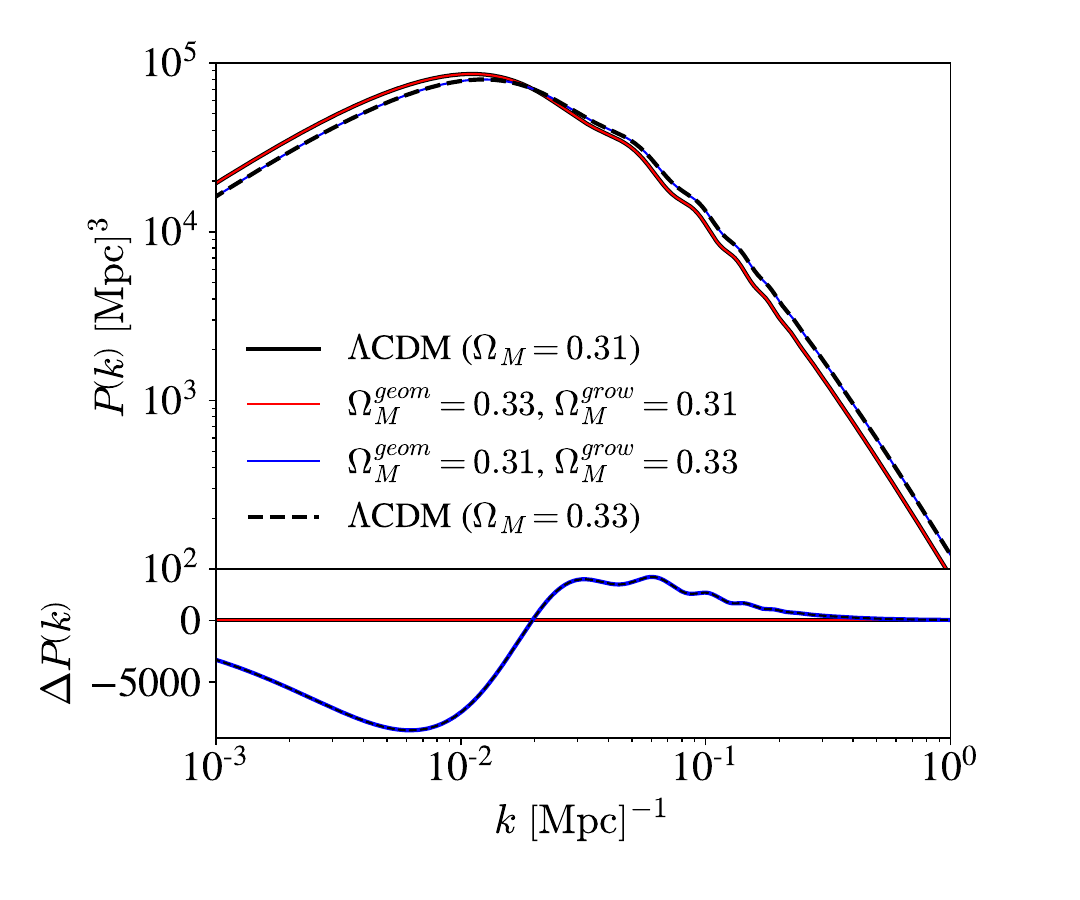}
    \caption{Same as Fig.~\ref{fig:cmb_tt_split}, but for the matter power spectrum (once again at $z = 0$).  Note that the power spectrum is sensitive only to the growth sector; see text for details.}
    \label{fig:mpk_split}
\end{figure}

Figure \ref{fig:mpk_split} shows the change in the matter power spectrum at $z = 0$ due to varying $\omgeom$ and $\omgrow$. Note that only the growth parameter, $\omgrow$, induces a change. This is because the geometry parameters affect only the transformation from distance $\chi$ to the scale $k$. Thus, at a series of fixed $k$ values --- that is, in a $P\left(k\right)$ plot --- the coordinate is determined uniquely by the growth parameters. The geometry enters in the conversion from distance $\chi$ to wavenumber $k=\ell/\chi$.

Here we utilize the KiDS--1000 public data\footnote{\url{http://kids.strw.leidenuniv.nl/DR4/}}, which is described in detail in Refs.~\cite{Kuijken:2019gsa,Giblin:2020quj}. This includes the $\xi_+(\theta)$ and $\xi_-(\theta)$ data vectors for both the auto- and cross-correlations across four tomographic redshift bins. We employ the likelihood from Ref.~\cite{Asgari:2020wuj}, but modify it to ensure the input $\chi(z)$ is the distance computed using geometry parameters. We also adopt the same scale cuts in $\theta$ as KiDS \cite{Asgari:2020wuj}. 

We also follow the KiDS analysis in adopting COSEBIs (Complete Orthogonal Sets of E/B-Integrals~\citep{Schneider:2001af}) as our summary statistic. COSEBIs are defined from the two-point correlation function via \cite{Asgari:2020wuj}
\begin{equation}
\begin{aligned}
\label{eq:COSEBIsReal}
 E_n &= \frac{1}{2} \int_{\theta_{\rm min}}^{\theta_{\rm max}}
 d\theta\,\theta\: 
 [T_{+n}(\theta)\,\xi_+(\theta) +
 T_{-n}(\theta)\,\xi_-(\theta)]\;, \\[0.2cm]
 B_n &= \frac{1}{2} \int_{\theta_{\rm min}}^{\theta_{\rm
     max}}d\theta\,\theta\: 
 [T_{+n}(\theta)\,\xi_+(\theta) -
 T_{-n}(\theta)\,\xi_-(\theta)]\;,
\end{aligned}
\end{equation}
where $T_{\pm n}(\theta)$ are filter functions defined for a given angular range, i.e.\ between $\theta_{\rm min}$ and $\theta_{\rm max}$. The logarithm of COSEBIs defined in Eq.~(\ref{eq:COSEBIsReal}) provides efficient data compression, with just a few COSEBI modes encoding most of the information in the measurements; here the index, $n$, varies over a small range of integers. We utilize the first five modes (so $1\leq n\leq 5$), following Ref.~\cite{Asgari:2020wuj}.
In general, COSEBIs provide the following benefits over the conventional two-point shear correlation functions: (i) they are less sensitive to baryonic feedback, and so, can capture information on smaller scales, and; (ii) since the weak lensing effect is sourced by a gradient of a gravitational potential, it cannot produce B modes, or curl modes. Hence, any detected B-modes in the COSEBI decomposition will arise from systematics and allow for a more accurate calibration of said systematics.

\subsection{BAO/RSD split}\label{sec:BAO}

Baryon acoustic oscillations (BAO) refer to the coherent oscillations that took place in the baryon-photon fluid in the epoch prior to recombination. The BAO imprint a characteristic scale in the distribution of matter in the universe, and this scale is given by the sound-horizon distance evaluated out to the baryon drag epoch 
\begin{equation}
    r_d = \frac{c}{\sqrt{3}} \int_0^{a_{\rm drag}}\frac{da}{a^2 H(a) \sqrt{1+\displaystyle\frac{3\Omega_B}{4\Omega_\gamma}}a} \,,
\end{equation}
where $\Omega_\gamma$ and $\Omega_B$ are the photon and baryon densities relative to the critical, respectively. The BAO feature can be observed with galaxy distributions as an excess probability for the clustering of galaxies separated by this characteristic scale. Redshift space distortions (RSD), on the other hand, refer to specific anisotropic features in the clustering of galaxies on small scales due to the impact of large-scale structures on the galaxies' velocities.

BAO/RSD surveys measure galaxy and quasar clustering, and are thus nominally represented by the anisotropic power-spectrum measurements $P(k, z, \mu)$, where $\mu$ is the cosine of the wavenumber direction to the line-of-sight. Most often, however, those measurements are compressed into a few, simpler meta-quantities. These compressed quantities are motivated by what the BAO effectively measure, which is the the angular feature on the sky given by either $D_A=r_d/\theta$ for angular clustering or $H(z)=c \Delta z / r_d $ for radial clustering; here $\theta$ and $\Delta z$ are, respectively the angle and redshift at which the excess of galaxy clustering due to the BAO is observed. If a survey can successfully separate information from transverse (angular) and radial modes, then it typically reports the two corresponding distance measures
\begin{eqnarray}
    D_{M}\left(z\right)&\equiv&\frac{c}{H_{0}}\int_{0}^{z}dz'\frac{H_{0}}{H\left(z'\right)}\,, \label{DM}\\[0.2cm]  
    D_{H}\left(z\right)&\equiv&\frac{c}{H\left(z\right)}, \label{DH}   
\end{eqnarray}
while surveys that do not attempt to separate the different modes just report a single, volume-averaged distance
\begin{equation}
    D_{V}\left(z\right) \equiv \left[zD_M^2\left(z\right)D_H\left(z\right)\right]^{1/3}.
    \label{DV}
\end{equation}
Since the basic quantity that the BAO measures is an angle (subtended by the sound-horizon standard ruler), all of these distance measurements are typically reported in units of $r_d$. The latter quantity is evaluated in the standard \LCDM\ model and held constant in the analysis.

In addition, spectroscopic surveys also measure redshift-space distortions (RSD): anisotropic features in galaxy clustering which are sensitive to the quantity $f\sigma_8$, where $f(a)\equiv d\ln D/d\ln a$ is the linear growth rate with $D(a)$ being the linear growth, and $\sigma_8$ is the amplitude of mass fluctuations on scales $8\hinvmpc$. The dimensionless quantity $f\sigma_8$ is an excellent probe of dark-energy models, and is determined purely by the growth of cosmic structure.

In this work, we use BAO/RSD data\footnote{ \url{https://svn.sdss.org/public/data/eboss/DR16cosmo/tags/v1_0_0/likelihoods/BAO-plus/}} from the following datasets: SDSS-DR7 Main Galaxy Sample~\cite{Ross:2014qpa}, BOSS-DR12 LRG~\cite{Alam:2016hwk}, eBOSS-DR16 LRG, eBOSS-DR16 QSO, and eBOSS-DR16 Ly$\alpha$ auto- and cross-correlation with QSO~\cite{Alam:2020sor}. These can be treated as mutually independent datasets, but we make sure to take into account the provided covariance matrices between the different measurements inside the \textit{same} catalog. The measurements that we adopt are presented in Tab.~\ref{tab:bao}. Note that the Ly$\alpha$ likelihoods are not well approximated by gaussians, so, in these cases we have made use of a grid of the probabilities as function of $D_M/r_d$, and $D_H/r_d$. 

% vertical and hor stretch, respectively
\renewcommand{\arraystretch}{1.4}
\setlength\tabcolsep{0.2cm}
\begin{table*}[]
\centering
\begin{tabular}{l|c|c|c|c|l}
\hline\hline
Catalog                                        & quantity      & $z$                      & measurement                 & $\sigma$                & Ref.                                                   \\ \hline
\multirow{2}{*}{BOSS-DR7 MGS}                  & $D_{V}/r_{d}$ & \multirow{2}{*}{$0.15$}  & $4.51$                      & $0.16$                  & \multirow{2}{*}{\cite{Ross:2014qpa}}  \\ \cline{2-2} \cline{4-5}
                                               & $f\sigma_{8}$ &                          & $0.53$                      & $0.14$                  &                                                        \\ \hline
\multirow{6}{*}{BOSS-DR12 LRG}                 & $D_{M}/r_{d}$ & \multirow{3}{*}{$0.38$}  & $10.27$                     & $0.15$                  & \multirow{6}{*}{\cite{Alam:2016hwk}}  \\ \cline{2-2} \cline{4-5}
                                               & $D_{H}/r_{d}$ &                          & $24.89$                     & $0.58$                  &                                                        \\ \cline{2-2} \cline{4-5}
                                               & $f\sigma_{8}$ &                          & $0.497$                     & $0.045$                 &                                                        \\ \cline{2-5}
                                               & $D_{M}/r_{d}$ & \multirow{3}{*}{$0.51$}  & $13.38$                     & $0.18$                  &                                                        \\ \cline{2-2} \cline{4-5}
                                               & $D_{H}/r_{d}$ &                          & $22.43$                     & $0.48$                  &                                                        \\ \cline{2-2} \cline{4-5}
                                               & $f\sigma_{8}$ &                          & $0.459$                     & $0.038$                 &                                                        \\ \hline
\multirow{3}{*}{eBOSS-DR16 LRG}                & $D_{M}/r_{d}$ & \multirow{3}{*}{$0.698$} & $17.65$                     & $0.30$                  & \multirow{10}{*}{\cite{Alam:2020sor}} \\ \cline{2-2} \cline{4-5}
                                               & $D_{H}/r_{d}$ &                          & $19.77$                     & $0.47$                  &                                                        \\ \cline{2-2} \cline{4-5}
                                               & $f\sigma_{8}$ &                          & $0.473$                     & $0.044$                 &                                                        \\ \cline{1-5}
\multirow{3}{*}{eBOSS-DR16 QSO}                & $D_{M}/r_{d}$ & \multirow{3}{*}{$1.48$}  & $30.21$                     & $0.79$                  &                                                        \\ \cline{2-2} \cline{4-5}
                                               & $D_{H}/r_{d}$ &                          & $13.23$                     & $0.47$                  &                                                        \\ \cline{2-2} \cline{4-5}
                                               & $f\sigma_{8}$ &                          & $0.462$                     & $0.045$                 &                                                        \\ \cline{1-5}
\multirow{2}{*}{eBOSS-DR16 Ly$\alpha$ (auto)}  & $D_{M}/r_{d}$ & \multirow{4}{*}{$2.334$} & \multicolumn{2}{c|}{\multirow{4}{*}{grid likelihood}} &                                                        \\ \cline{2-2}
                                               & $D_{H}/r_{d}$ &                          & \multicolumn{2}{c|}{}                                 &                                                        \\ \cline{1-2}
\multirow{2}{*}{eBOSS-DR16 Ly$\alpha$ (cross)} & $D_{M}/r_{d}$ &                          & \multicolumn{2}{c|}{}                                 &                                                        \\ \cline{2-2}
                                               & $D_{H}/r_{d}$ &                          & \multicolumn{2}{c|}{}                                 &                                                        \\ \hline\hline
\end{tabular}
\caption{Summary of the BAO and RSD data used in this analysis. The columns show, from left to right, the name of the dataset, the quantity in question, mean redshift at which the quantity is measured, the measurement, its error, and reference from which the measurement was adopted. The ``grid likelihood'' denotes quantities for which a grid likelihood (as opposed to a single Gaussian measurement with an error) was provided. Note that we used the full covariance matrix to combine all measurements.} 
\label{tab:bao}
\end{table*}

Our choices for the geometry-growth split in BAO/RSD measurements are as follows:
\begin{itemize}
    \item The drag horizon $r_d$ is calculated using the growth parameters. This is consistent with our classification of all pre-recombination physics as being sensitive to only growth (e.g.\ the source function $S_T$ in the CMB anisotropy; see Sec.~\ref{sec:CMB})
    \item The distance $D_A$ and the Hubble parameter $H$ are defined as geometry in our split as they fundamentally come from the measurements of the angular and radial separations.  
    \item  Finally, both quantities in the product $f\sigma_8$ are treated as growth as the amplitude and growth of structure is what causes the redshift-space distortions in the first place; see \cite{Ruiz:2014hma}.
\end{itemize}

Therefore, the combination of BAO and RSD measurements constitute a hybrid probe, containing information on both geometry (through $D_A$ and $D_H$) and growth (through $r_{d}$ and $f\sigma_{8}$).

\subsection{SN Ia split}

We adopt the Pantheon set of 1048 type Ia supernovae covering the redshift range
$0.01<z<2.26$ \citep{Scolnic:2017caz}. The apparent magnitude can be related to luminosity distance via
\begin{equation} \label{mu}
    m\left(z\right )=5\log_{10}\left[H_0D_L(z)\right ]+\mathcal{M}\,,
\end{equation}
where $\mathcal{M}$ is the nuisance parameter that combines the absolute distance of supernovae and the Hubble constant, and one that needs to be marginalized over. We use the full covariance matrix of the supernova magnitude measurements\footnote{\url{https://github.com/dscolnic/Pantheon}}, which consists of both signal and noise. 

Type Ia supernovae are used as a cosmological distance indicator, so it is natural to classify them as pure geometry. Therefore, we feed only the geometry parameters when computing the luminosity distance from theory. This choice agrees with all previous geometry-growth split work.

\subsection{Summary of the split choices and comparison to other works}

Table \ref{tab:summary} summarizes our choices for which components of the cosmological probes are sensitive to geometry vs.\ growth. We again emphasize that while some classifications of the probes are obvious, others are more subjective \cite{Wang:2007fsa}.
To illustrate how our choices compare to those made in the geometry-growth literature, we now compare our approach with those of previous works  \citep{Wang:2007fsa, Ruiz:2014hma, Bernal:2015zom, Muir:2020puy}. 

\begin{table}[!t]
\begin{center}
\begin{tabular}{l c c}
\hline \hline
Cosmological Probe & Geometry                                       & Growth \\ \hline 
SN Ia              & $H_0 D_L(z)$                                     &  ----- \\[0.2cm]
BAO                & $\ds\{D_M(z); D_H(z)\}$ &  $r_d(z_d)$ \\[0.35cm]
CMB\     &   $j_{\ell}[k \chi(z)]$      &  $S_T(k, z)$  \\[0.2cm]                

Weak lensing      & $\ds\frac{d\chi(z)}{d(z)} g_i(z) g_j(z)$         & \,\, $\Omega_m^2 P_{\delta}\left(\frac{\ell}{\chi}, z \right)$\,\, \\[0.35cm]
RSD                & -----                        &  $f(z)\sigma_8(z)$\\
\hline\hline
\end{tabular}
\end{center}
\caption{Summary of the cosmological probes used in this work and the related theoretical quantities for each that constrains either the geometry or the growth parameters. Here $r_d(z_d)$ refers to the sound horizon evaluated at the baryon drag epoch $z_d$. See text for more details.}
\label{tab:summary}
\end{table}

For the CMB information,  Ref.~\cite{Wang:2007fsa} classifies the primordial CMB fluctuations as growth, while the projection onto the 2D observed sky is determined by geometry. The result is that the angular power spectrum is sensitive to both geometry and growth parameters. Ref.~\cite{Ruiz:2014hma}, on the other hand, only uses the compressed CMB information present in the CMB acoustic peaks’ locations, and therefore considered CMB as a geometry-only probe. Recently, the Dark Energy Survey (DES) collaboration \cite{Muir:2020puy} followed the same approach, arguing that using the compressed CMB information --- which provides only geometric information --- reflects the fact that the CMB is mostly sensitive to mapping out the angular scale of the sound horizon. Finally, Ref.~\cite{Bernal:2015zom} chose the high multipoles in the angular power spectrum to constrain geometry while the low multipoles constrained growth. Additionally, they chose the lensing power spectrum in the multipole range $40 \leq L \leq 400$ to be growth (see their Table 1 for details). Our choice, on the other hand, is motivated by the desire to extract the full information from the ``building blocks'' of the CMB observations: the primordial fluctuations and their projection. In doing so, our geometry-growth split most closely follows that of Ref.~\cite{Wang:2007fsa}, with the corresponding behavior from the split parameterization illustrated in Figure \ref{fig:cmb_tt_split}.

For weak lensing, Refs.~\citep{Wang:2007fsa} and \cite{Ruiz:2014hma} have similar strategies. They, however, differ in how they treat the $\Omega_M^2$ prefactor (see Eq.~(\ref{eq:convergence_z})); the former paper includes this as a growth quantity, while the latter paper considers it a part of the lensing window function, and hence a geometric quantity. In the present work, we opt to treat the $\Omega_M^2$ factor as a growth parameter, as  shown in Table \ref{tab:summary}. Ref.~\cite{Bernal:2015zom} has not used weak leasing at all, citing difficulty in disentangling growth and geometry contributions for this probe. The DES \cite{Muir:2020puy} roughly follows the implementation from \citep{Ruiz:2014hma}, with an additional modification in modelling the matter power spectrum. The DES implement a redshift-dependent split in the linear matter power spectrum; at $z>3.5$, their $P_{\rm lin}(k, z)$ is given by just geometry parameters (in concert with treatment of the CMB as geometry only), while at $z<3.5$ it is given by taking the matter power spectrum at $z=3.5$ and scaling it by the squared ratio of the growth functions since that redshift, where the latter quantity is computed with growth parameters alone. 

The DES compute the non-linear matter power spectrum using halofit on the modified linear power spectrum function, $P_{\rm lin}(z,k)$, but pass in only growth parameters to the halofit method. They also utilize growth parameters in the intrinsic alignment model for the lensing predictions. Our choice, in contrast, is to model the matter power spectrum as pure growth (see Figure \ref{fig:mpk_split}). We emphasize that approaches in all of the aforementioned papers are completely self-consistent. 

For BAO, the usual classification found in the literature is to consider it as a purely geometrical probe \citep{Ruiz:2014hma, Bernal:2015zom, Muir:2020puy}. We, however, implement a different strategy, and argue that while the distance and the Hubble parameter in BAO are still geometric quantities, the sound horizon that enters the compressed ``observable'' quantities (see Tab.~\ref{tab:bao} and Eq.~\eqref{DV}) should be computed using growth parameters. The latter choice, while perhaps also subjective, is fully consistent with our treatment of the sound horizon for the CMB, and also makes the BAO in our analysis sensitive to both geometry and growth.
 
%=================================================================
\begin{figure*}[t!]
\centering
\includegraphics[width=0.325\textwidth]{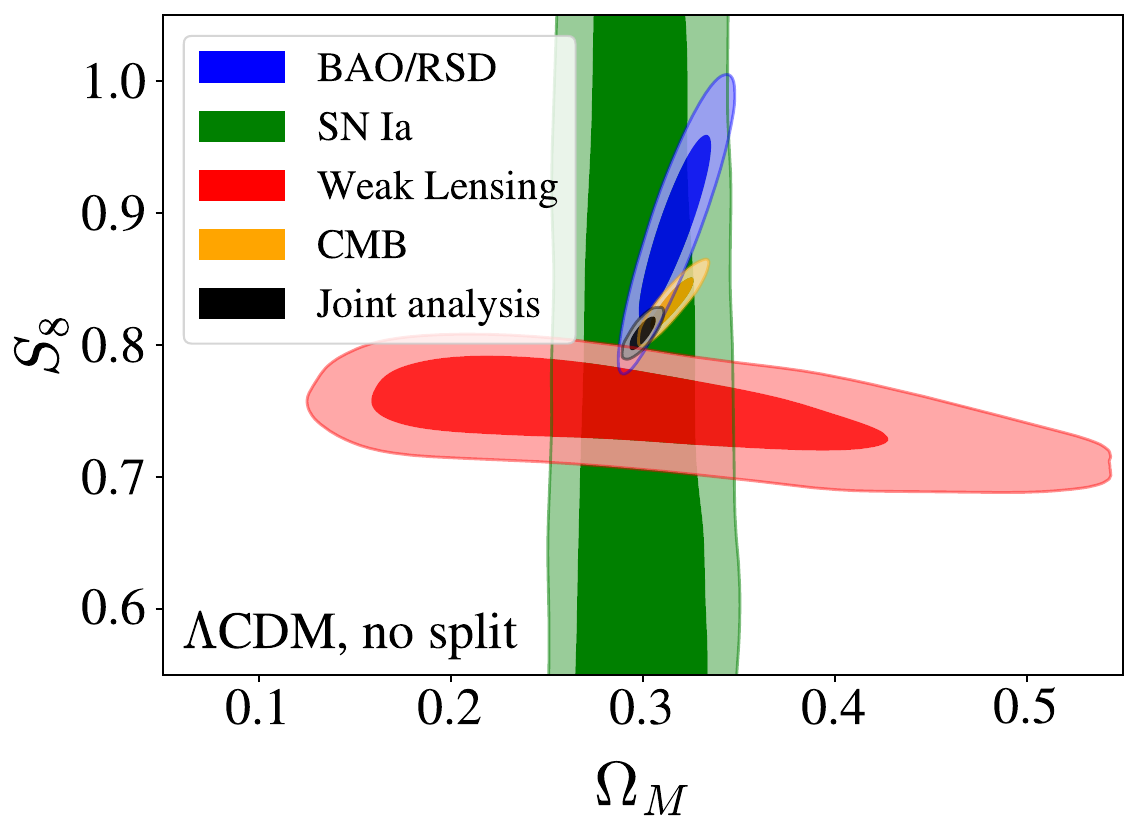}
\includegraphics[width=0.325\textwidth]{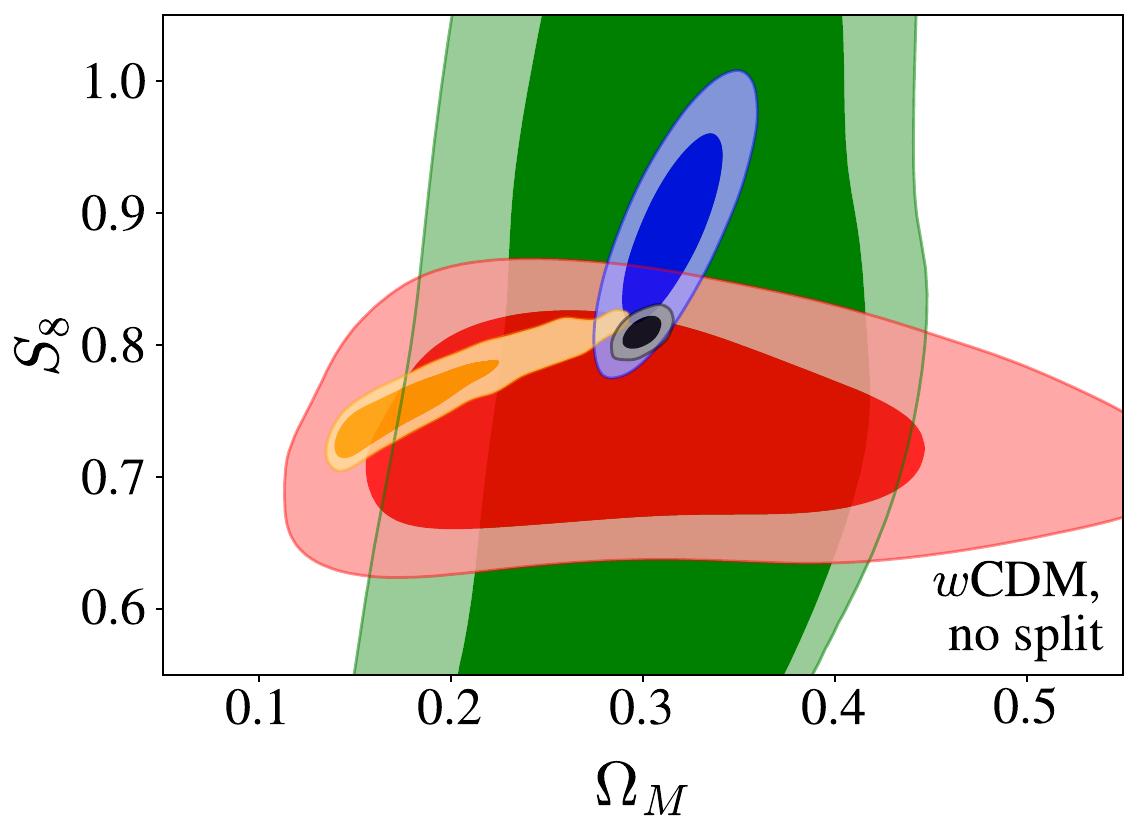}
\includegraphics[width=0.325\textwidth]{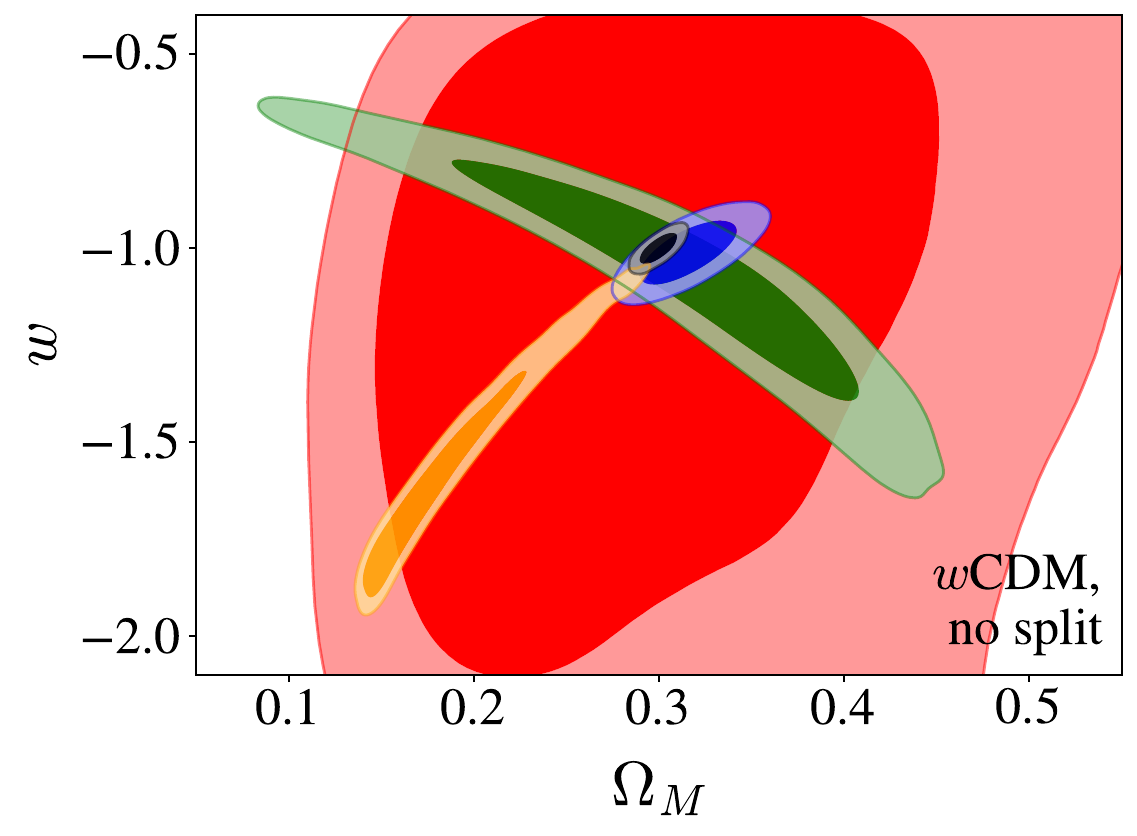}
\caption{Marginalized 2D posteriors, of individual probes as well as the joint analysis, for the fiducial models where no geometry-growth split has been implemented. \textit{Left panel:} $\Omega_{M}$-$S_{8}$ plane in the \LCDM\ model.  \textit{Middle panel:} $\Omega_M$-$S_8$ plane in the wCDM model. \textit{Right panel:} $\Omega_M-w$ plane in the wCDM model.}
\label{fig:LCDM_wCDM}
\end{figure*}
%=================================================================

The RSD information is usually compressed into measurements of $f(z)\sigma_8(z)$ at several redshifts. Here, we follow the same procedure as in Refs.~\citep{Wang:2007fsa, Ruiz:2014hma, Bernal:2015zom}, where $f\sigma_8$ depends only on the growth parameters. The only work that differs from this thus far is Ref.~\cite{Muir:2020puy} who allowed $\sigma_8(z=0)$ to also include geometric parameters via their split parameterization of the matter power spectrum, $P_{lin}^{split}(k,z)$ (see their Sec.~II D for details). 

Finally, type Ia supernovae measure relative values of the luminosity distance, and so all works --- including this one --- assume that SN Ia constrain only the geometry parameters.

\section{Results}\label{sec:results}

We now present our main results --- constraints on the geometry and growth parameters $(\omgeom, \omgrow)$ in the extended \LCDM\ model, and on $(\omgeom, \omgrow, \wgeom, \wgrow)$ in the extended wCDM model. For all marginalized 2D posteriors we show the 68\% and 95\% contours.

\subsection{Analysis setup}

Our base set of cosmological parameters is given by Eq.~\eqref{paramsom} when we do the \lcdm\ split, and by Eq.~\eqref{paramsomw} when we consider the \wcdm\ split. In both cases the set of parameters that are \textit{not} being split, $\{p_i\}$, is
\begin{equation} \label{params}
    \{p_i\}=\left\lbrace\omega_{b}, H_{0}, \ln (10^{10}A_s), n_{s}, \tau_{\rm reio}\right\rbrace \,,
\end{equation}
where $\omega_b\equiv\Omega_b h^{2}$ is the physical baryon density, $H_0$ is the Hubble constant, $A_{s}$ is the amplitude of the primordial power spectrum at $k_{\rm piv}=0.05\ {\rm Mpc}^{-1}$, $n_s$ is the scalar spectral index, and $\tau_{\rm reio}$ is the optical depth to reionization. As mentioned briefly in Sec.~\ref{sec:CMB}, in practice we vary the CDM density parameter $\Omega_c$ (rather than $\Omega_M$), so that $ \omgeom\equiv \Omega_c^{\rm geom} + \Omega_b$ for geometry, and similarly for growth.

We adopt flat priors in all base parameters, as shown in Tab.~\ref{tab.priors}. Note that the amplitude of mass fluctuations $\sigma_8$ is a derived parameter, as is the parameter
\begin{equation}
S_8\equiv \sigma_8 
\left (\frac{\Omega_M}{0.3}\right )^{0.5}.
\end{equation}
In the split models, $S_8$ is defined using only $\omgrow$.
The cosmological probes we employ in our analysis are presented in Tab.~\ref{tab:summary} and discussed in Sec.~\ref{sec:split_def}.

\begin{table}[h!]
\centering
\begin{tabular}{l|l}
\hline \hline
Parameter                                   & Flat Prior        \\ \hline
$100\ \omega_{b}$                           & $[1.875,2.625]$ \\
$H_{0}$                                     & $[60,80]$       \\
$\ln (10^{10}A_s)$                          & $[1.7,5]$       \\
$n_{s}$                                     & $[0.7,1.3]$     \\
$\tau_{\rm reio}$                               & $[0.004,0.1]$   \\
$\omgeom, \omgrow$     & $[0.01,0.99]$   \\
$w^{\rm geom}, w^{\rm grow}$                       & $[-3,1]$        \\ \hline \hline
\end{tabular}
\caption{Cosmological parameters and their respective flat priors used in the parameter selection analysis.}
\label{tab.priors}
\end{table}

To obtain the constraints, we use a suitably modified version of the Boltzmann code {\sc CLASS}~\cite{Blas:2011rf} alongside the MCMC code {\sc MontePython}~\cite{Audren:2012wb,Brinckmann:2018cvx}. In practice, the implementation of the split approach in the {\sc CLASS} code does not induce any appreciable loss in computational time needed to calculate the cosmological observables. On the other hand, since the split technique introduces new parameters and, consequently, new degeneracies, the computational time spent on MCMC sampling is considerably longer. For example, in the joint analysis --- where we combine all available probes --- the split case takes three or four times longer than the unsplit case.

We explore the parameter space using the Metropolis-Hastings sampler, and stop upon convergence, which is indicated by the Gelman-Rubin criterion ~\cite{Gelman:1992zz} of $R - 1 < 0.05$. This threshold is similar to those considered in previous works~\cite{Ruiz:2014hma,Bernal:2015zom}. Figures and analysis of the resulting chains are performed using the {\sc GetDist}~\cite{Lewis:2019xzd} package. 

%=================================================================
\begin{figure*}[t]
\centering
\includegraphics[width=0.49\textwidth]{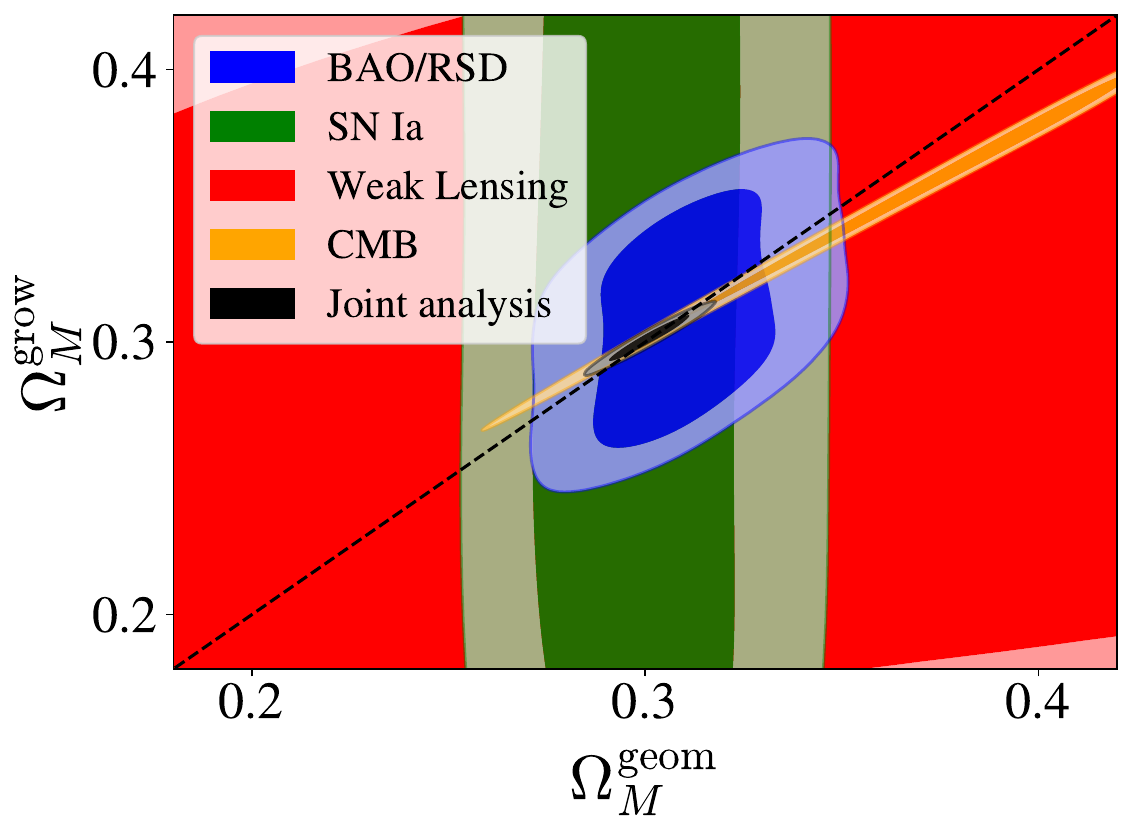}
\includegraphics[width=0.49\textwidth]{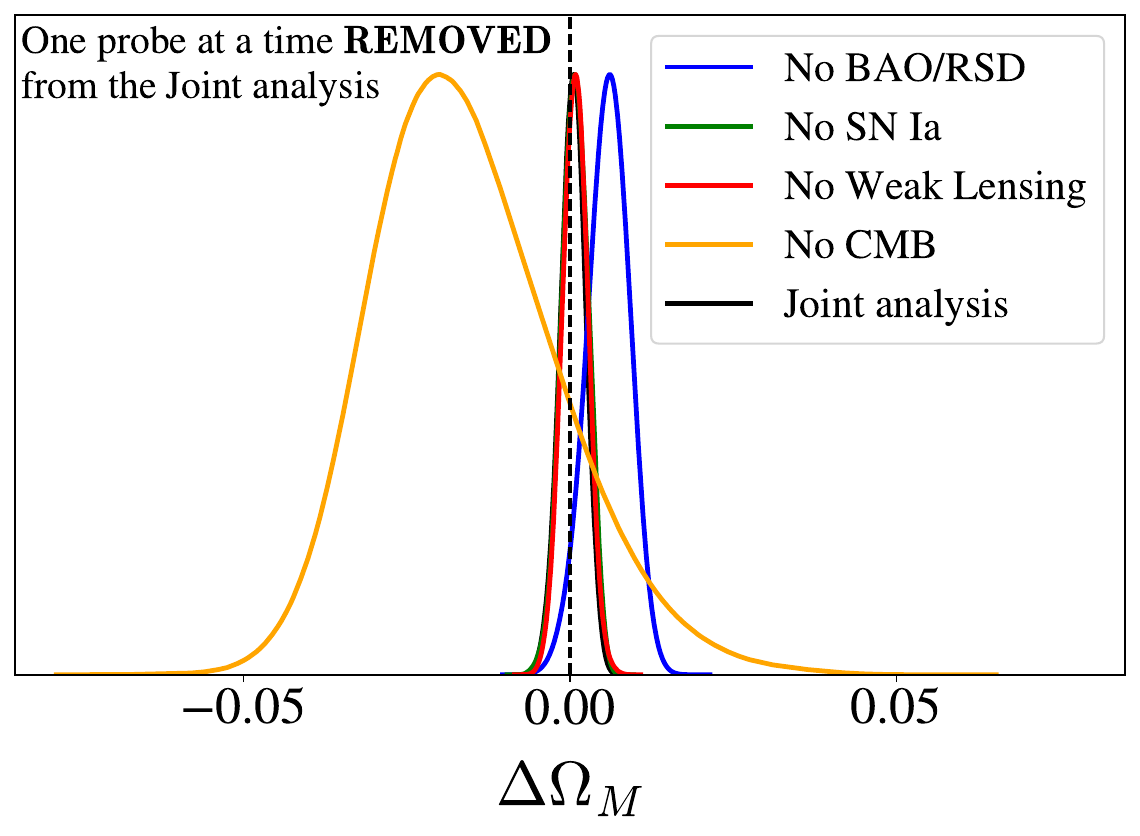}
\caption{Geometry-growth constraints for all individual probes as well as the joint analysis in the split \lcdm\ model, where the matter density which has been split into $\omgeom$ and $\omgrow$.  \textit{Left panel:} $\omgeom-\omgrow$ plane. The dashed black line shows the equality limit of $\omgeom = \omgrow$. \textit{Right panel:} One-dimensional posteriors for the difference $\dom=\omgrow-\omgeom$, which is particularly well-constrained by the CMB. The vertical dashed black line shows $\dom = 0$. Note that the right panel shows constraints where one probe at a time has been \textit{removed} from the joint analysis.}
\label{fig:om_split}
\end{figure*}

\subsection{Unsplit case}

We first analyze the standard cosmological model without the geometry-growth split. This is  helpful in order to 1) confirm consistency with similar results in the literature, and 2) provide a fiducial reference for comparisons with the split results that follow below. 

In line with the split analysis to follow, we present results for the two basic cosmological models: flat \LCDM, and flat wCDM.  Whereas in the first case we have the usual six standard free cosmological parameters (equivalent to Eqs.~\eqref{paramsom} and~\eqref{params} reduced by the condition $\omgeom=\omgrow\equiv\Omega_{M}$), in the second case we have seven parameters, corresponding to Eqs.~\eqref{paramsomw} and~\eqref{params} with the conditions $\omgeom=\omgrow\equiv\Omega_{M}$ and $\wgeom=\wgrow\equiv w$. Our results for both unsplit models are shown in Fig.~\ref{fig:LCDM_wCDM}.

The left panel of  Fig.~\ref{fig:LCDM_wCDM} shows the $\Omega_M$-$S_8$ contour in the LCDM model. As expected, SN Ia are insensitive to the parameter $S_8$ which largely encodes the amplitude of mass fluctuations. Weak lensing, on the other hand, is sensitive to both $\Omega_M$ and $S_8$, and places stronger constraints on $S_8$ than on $\Omega_M$. The BAO/RSD data also constrains both parameters; its dependence on $S_8$ comes exclusively from the RSD quantity $f\sigma_8$ (see Tab.~\ref{tab:summary}). The CMB provides the best individual constraints in this plane. The combined contour is very small in comparison to the others due to degeneracy breaking in this multi-dimensional parameter space --- something that will become even more accentuated once we go to the split analyses. Focusing on the KiDS and CMB analyses, it is worth mentioning that we reproduce the aforementioned $S_{8}$ tension (e.g.\ ~\cite{DiValentino:2020vvd}). From this combined analysis the constraint on the matter density that we obtain is
\begin{equation}
     \Omega_M = 0.2998_{-0.0047}^{+0.0044}\qquad\text{(\LCDM, no split)}.
\end{equation}
The constraints for the other parameters are provided in Tab.~\ref{tab.results}. While there exist prospective inconsistencies between different probes (e.g., $S_8$ tension between KiDS-1000 and Planck 2018) we proceed to combine all data in a joint analysis since our main focus is not fiducial parameter constraints, but to establish a reference result that can be compared with constraints from the split cases.

 Fig.~\ref{fig:LCDM_wCDM} show the $\Omega_M$-$S_8$ contour (middle panel) and $\Omega_M$-$w$ contour (right panel) of our the unsplit wCDM analysis. The former plot is similar to the one for \LCDM\ case, except that the contours --- especially for BAO/RSD and CMB ---  are broadened due to the additional degeneracy with the dark energy equation of state parameter. The $\Omega_M$-$w$ plot shows impressive complementarity of the different cosmological probes in breaking the degenaricies in this plane; a feature first pointed out two decades ago \citep{Huterer:1998qv}. The figure also shows that most of these probes, and especially weak lensing, do not give strong constraints on their own, yet play a critical role by breaking degeneracies in the joint analysis, as indicated by the remarkably small contour for the joint analysis. The combined constraints on $\Omega_M$ and $w$ are
\begin{equation}
  \begin{aligned}  
 \Omega_M &= 0.2997_{-0.0066}^{+0.0066} 
 \\[0.2cm] 
 w &= -1.002_{-0.024}^{+0.027}
 \end{aligned}
 \qquad\text{(wCDM, no split)}.
\end{equation}
Once again, constraints for all other parameters can be found in Table~\ref{tab.results}.

\begin{figure*}[]
\centering
\includegraphics[width=0.48\textwidth]{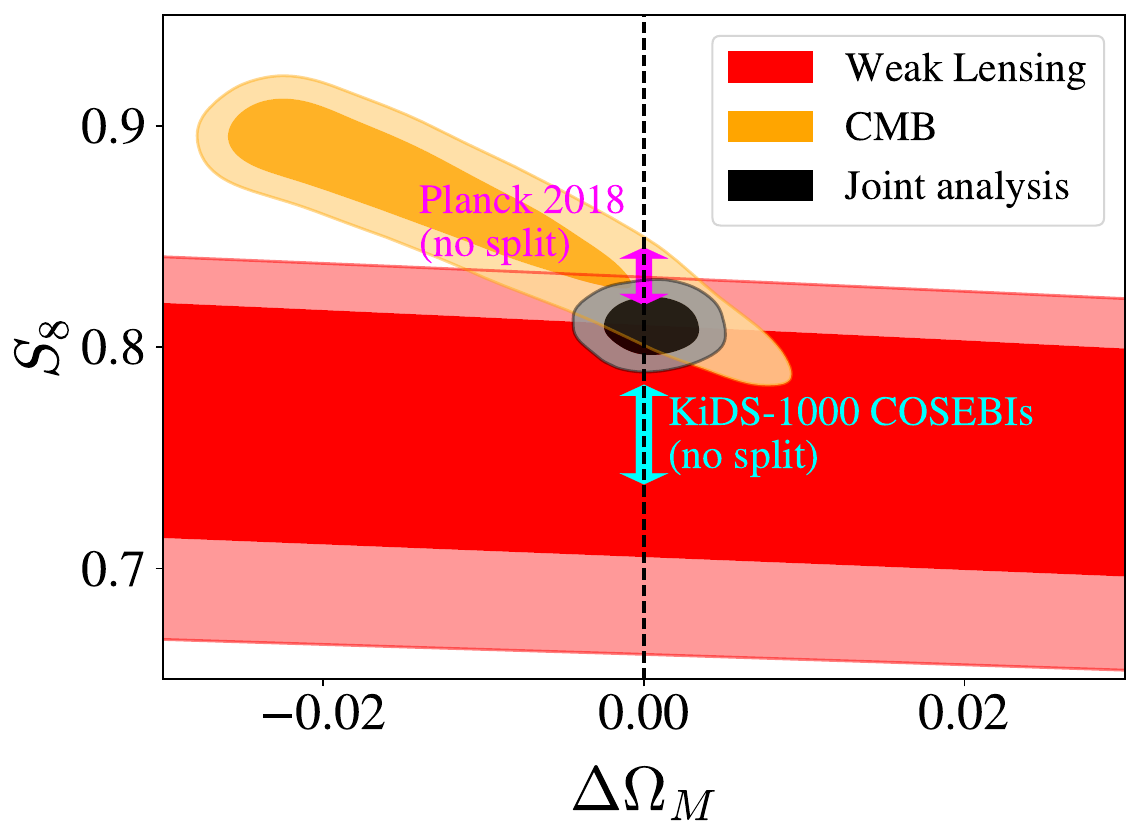}
\includegraphics[width=0.5\textwidth]{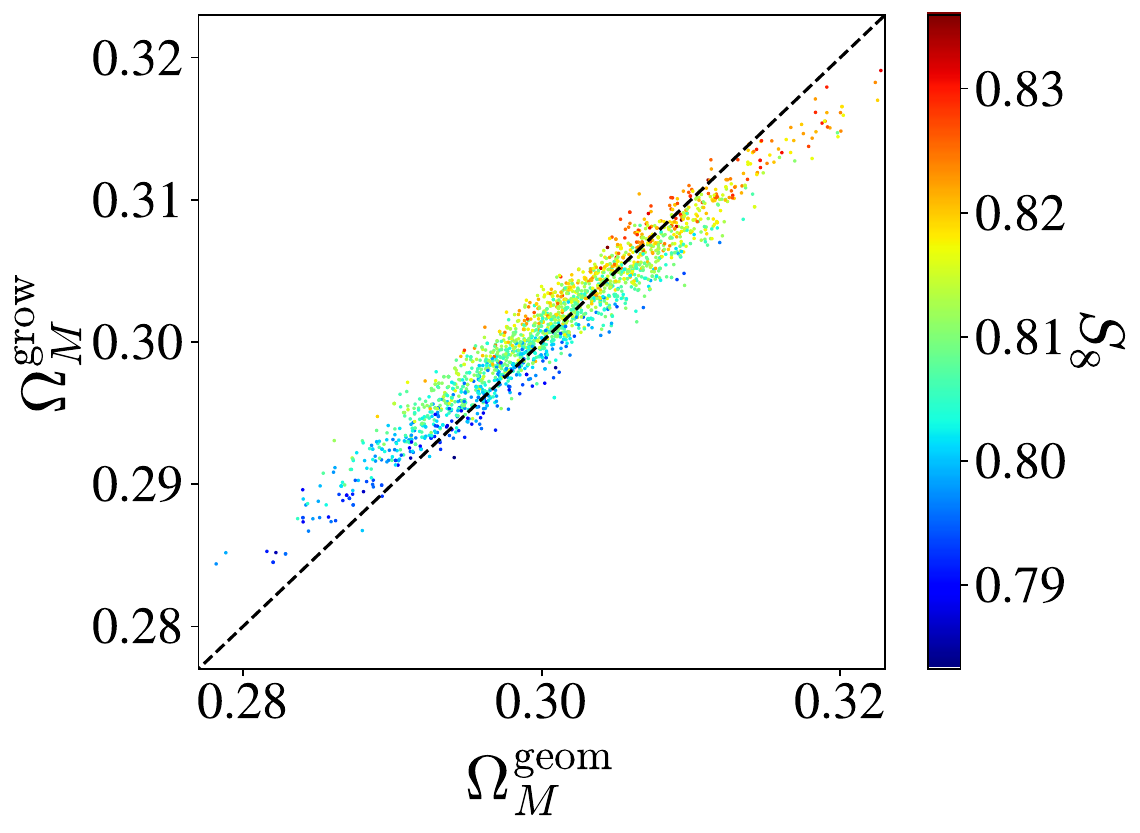}
\caption{\textit{Left panel:} The marginalized posteriors of $\dom\equiv\omgrow-\omgeom$ and \ $S_8$ in the split \LCDM\ case, for CMB, weak lensing, and the joint analysis (which includes BAO, RSD, and SN Ia as well). The $S_8$ constraints ($1\sigma$) from Planck 2018 (TTTEEE+lensing)~\cite{Aghanim:2018eyx}, and KiDS-1000 COSEBIs~\cite{Asgari:2020wuj} are shown as the magenta and cyan arrows, respectively. The parameter $\dom$ is correlated with $S_8$ for both probes, with the CMB showing a slightly stronger correlation. The vertical black dashed line shows $\dom = 0$. \textit{Right panel:} Points from the posterior in the $\omgeom - \omgrow$ plane, color-coded by the value of $S_{8}$ in the joint analysis of the \lcdm\ split model.}
\label{fig:om_split1}
\end{figure*}

\subsection{Split \LCDM\ model}\label{sec:split_LCDM}

Moving to the cases where the split parameterization is adopted, we first consider the split \LCDM\ model. Here the matter density, $\Omega_M$, has been promoted into two parameters responsible for geometry and growth, respectively. The cosmological parameter set is given by Eqs.~\eqref{paramsom} and~\eqref{params}. Key results are  shown in Figs.~\ref{fig:om_split} and \ref{fig:om_split1}, and in more detail in Fig.~\ref{fig:omctriangle} (in the Appendix) and in Tab.~\ref{tab.results}. 

The constraints from individual probes on familiar parameters such as $H_0$ and $S_8$ are significantly weaker than those in the standard, unsplit \LCDM\ model (see the Appendix~\ref{appx:details}). However, combining all the probes to perform a joint analysis enables massive degeneracy breaking in the full parameter space \citep{Ruiz:2014hma}, leading to fairly strong constraints not only on the geometry sector, but even on the tough-to-constrain growth sector.

In the $\omgeom-\omgrow$ plane (left panel of Fig.~\ref{fig:om_split}) we find, as expected, that the SN Ia probe constrains $\omgeom$ but has no sensitivity to $\omgrow$. We also see that the CMB provides an excellent constraint on the \textit{difference} between the geometry and growth matter densities,

\begin{equation}
    \dom\equiv\omgrow-\omgeom.
    \label{eq:dom}
\end{equation}
This is entirely expected given our previous results from  Sec.~\ref{sec:CMB} and Fig.~\ref{fig:cmb_tt_split}. The CMB, however, constrains the \textit{sum} $\omgeom + \omgrow$ much more poorly. Fortunately, the SN Ia and BAO/RSD data play a crucial role in breaking the degeneracy in this latter combination. 

As mentioned before, the combination of all probes, which is our joint analysis, provides strong constraints on most cosmological parameters. The constraints we obtain on the matter density parameters are
\begin{equation}
  \begin{aligned}  
 \omgeom &= 0.3012_{-0.0071}^{+0.0066}
 \\[0.2cm] 
 \omgrow &= 0.3017_{-0.0056}^{+0.0057} 
 \end{aligned}
 \qquad\text{(split \LCDM)}.
\end{equation}
We see that the geometry and growth matter-density are in remarkably good agreement with each other, despite the small error bars in each. 
In general, the only appreciable differences between the split \lcdm\ constraints and those of the unsplit \lcdm\ model are the slightly larger error bars on $H_{0}$ in the former (see Tab.~\ref{tab.results}).

We also show the posteriors on $\dom$ (right panel, Fig.~\ref{fig:om_split}). Since some of the individual probes deliver weak constraints on this parameter, we do \textit{not} show these individual-probe constraints, but rather the joint analysis constraints after removing one particular probe (e.g.\ the ``No weak lensing'' result corresponds to simultaneous constraints from BAO/RSD, SN Ia, and CMB data). Removing weak lensing from the analysis (red line, Fig.~\ref{fig:om_split}) gives essentially the same constraint as the full, unchanged joint analysis, which confirms the expectation (from e.g.\ the left panel of the same Figure) that weak lensing data alone do not appreciably constrain $\dom$. Removing any one of the other probes shows qualitatively similar results with slightly weaker constraints. As expected, the CMB provides by far the strongest constraint on $\dom$, as evidenced by the significantly weaker constraints obtained when it is removed from the joint analysis (orange line, Fig.~\ref{fig:om_split}). The numerical values for each of these constraints is shown in Tab.~\ref{tab:noresults}.

Following the procedure adopted in Ref.~\cite{Ruiz:2014hma}, we can quantify the statistical significance of departing from $\dom=0$ by computing the fraction of the posterior that satisfy the condition $\dom\lessgtr0$, 
\begin{equation} \label{pvalue}
    p=\frac{\int_{\dom\lessgtr0}d\dom\ \mathcal{L}\left(\dom\right)}{\int d\dom\ \mathcal{L}\left(\dom\right)}\,.
\end{equation}
In practice, Eq.~\eqref{pvalue} gives us the $p$-value for $\dom\lessgtr0$. For example, in the ``No BAO/RSD'' analysis we found $p=0.0430$ for $\dom<0$, and in the ``No CMB'' analysis we found $p=0.1370$ for $\dom>0$. Since we are looking at a one-dimensional posterior, the $p$-value can be roughly converted into the number of standard deviations using the one-dimensional Gaussian approximation for the tails. From this, the p-values above correspond to $2.0\sigma$ and $1.5\sigma$ confidence level, respectively, of inconsistency with the \lcdm\ model. 

The combined constraint on the difference between the geometry and growth values of $\Omega_M$ is 
\begin{equation}
\dom = 0.0004_{-0.0020}^{+0.0020}
\label{eq:dom_constraint}
\end{equation}
Note that even though the CMB provides the best individual measurement of this parameter, its individual constraint on $\dom$ is a factor of four weaker than the combined constraint in Eq.~(\ref{eq:dom_constraint}). This is another lesson on the wonderful complementarity of cosmic probes --- even those that do not appear strong individually --- to precisely constrain parameters of interest. This combined constraint corresponds to $0.9\sigma$ confidence level of inconsistency between geometric and growth parameters, which means we cannot reject the null hypothesis, i.e., $\dom=0$. 

%=================================================================
\begin{figure*}[!t]
\centering
\includegraphics[width=0.49\textwidth]{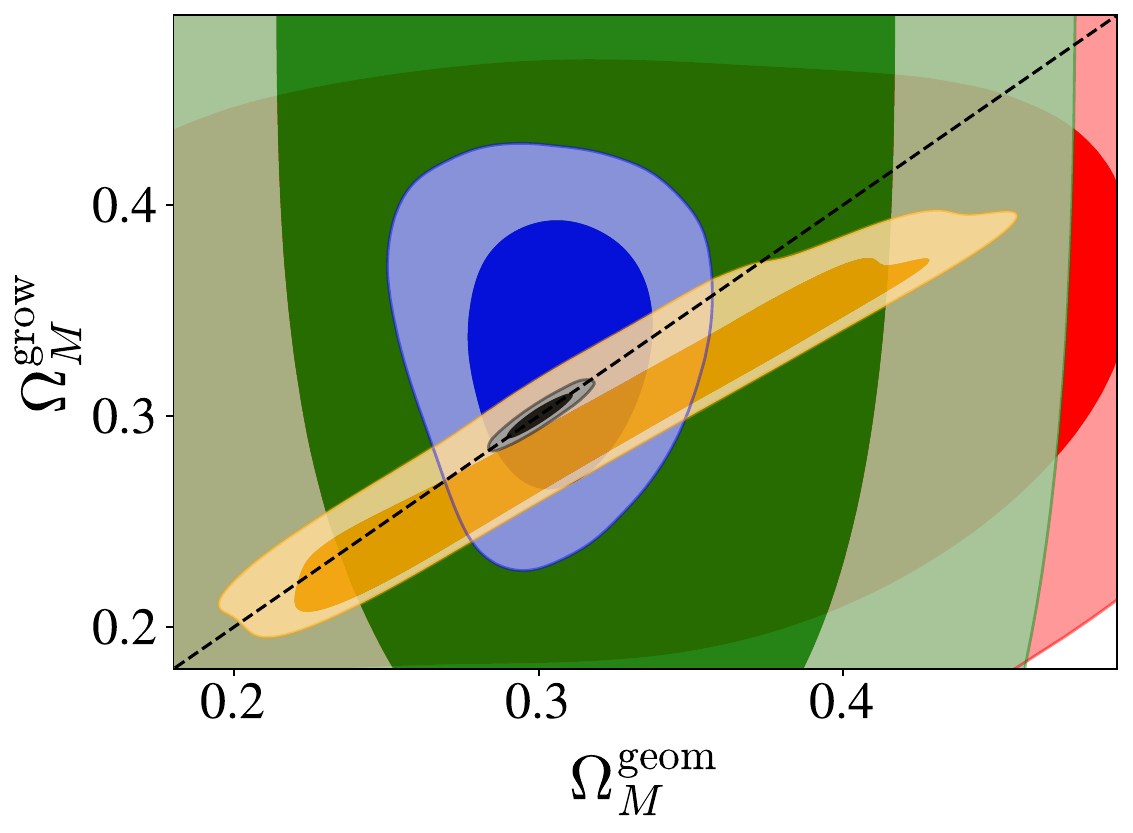}
\includegraphics[width=0.49\textwidth]{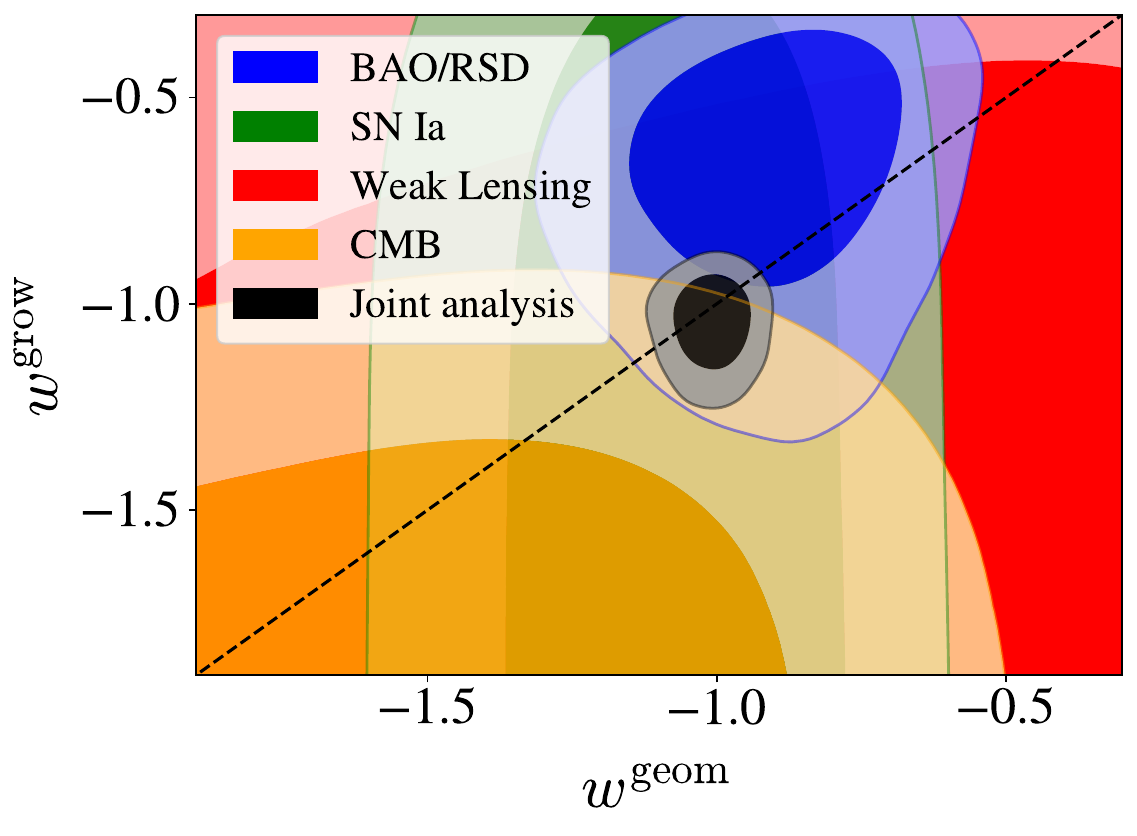}
\caption{Similar to Fig.~\ref{fig:om_split}, but for the split \wcdm\ model. We show constraints from the individual probes and the joint analysis for the $\omgeom-\omgrow$ (left) and $\wgeom-\wgrow$ (right) planes. The diagonal lines show the fiducial \wcdm\ model where $\omgeom=\omgrow$ and $\wgeom=\wgrow$, respectively. }
\label{fig:dom_dw}
\end{figure*}
%==================================================================

We are now well-positioned to assess whether the $S_{8}$ tension, observed in \LCDM, is \textit{correlated} with either geometry or growth. If so, it would be possible for the $S_8$ tension to be reinterpreted (and perhaps explained) by a difference between the geometry and growth parameters --- that is, a nonzero $\dom$. Such a conjecture would be supported, for example, by contours in the $\dom-S_8$ plane that show discrepant measurements of $S_8$ by CMB and weak lensing for $\dom=0$, but much better agreement between those two measurements for some $\dom\neq 0$.

Such a plot is shown in the left panel of Fig.~\ref{fig:om_split1}. Given our interest in the $S_{8}$ tension, we only show constraints from weak lensing, CMB, and the joint analysis (which uses all probes). The key message is that the weak lensing constraint on $\dom$ is very weakly correlated with $S_8$, and that result is likely to remain with any other current shear survey\footnote{For instance, the recent results released by the DES collaboration~\cite{Abbott:2021bzy,Secco:2021vhm} indicate that their constraints from the weak lensing analysis are of the same order as KiDS-1000 (see Fig.~10 of~\cite{Secco:2021vhm}).}. The CMB constraint on $\dom$ shows only a slightly steeper scaling with $S_8$, and so does not allow a sharp consistency check with weak lensing because the latter constraint is rather weak.
Therefore, we conclude that \textit{the quality of the present data does not allow a sharp test of whether a discrepancy between geometry and growth in the split-\lcdm\ model can explain the $S_8$ tension}. This conclusion is further supported by the right panel of Fig.~\ref{fig:om_split1} which shows that along lines of constant $\dom$ (roughly, going along the diagonal direction), $S_8$ can take a range of values, indicating that there is no discernable correlation between $S_8$ and the difference between $\omgeom$ and $\omgrow$.

\subsection{Split wCDM model}
\label{sec:split_wCDM}

We now consider the geometry-growth extensions of the flat \wcdm\ cosmological model, which includes a constant dark energy equation-of-state parameter, $w$. Here both $\Omega_M$ and $w$ have been split into their geometry and growth counterparts. The full cosmological parameter set is given by Eqs.~\eqref{paramsomw} and~\eqref{params}. Key results are shown in Fig.~\ref{fig:dom_dw}, and more detail is given in Fig~\ref{fig:omc-w-triangle} in the Appendix as well as Tab.~\ref{tab.results}. 

The constraints on $H_0$, $S_8$ and our four split parameters from \textit{individual} probes now show very significant degeneracies, as expected in this challenging-to-constrain parameter space; see the Appendix. However, when the probes are combined, we are again able to obtain fairly accurate constraints on most cosmological parameters. The degeneracy breaking power from using a diverse set of probes is once again at display here.

We focus on the split matter density and dark-energy equation of state in Fig.~\ref{fig:dom_dw}. The left panel shows the constraints from individual probes and the joint analysis in the $\omgeom-\omgrow$ plane, and the right panel is the same but for the $\wgeom-\wgrow$ plane.

As before, SN Ia are sensitive only to geometry parameters, while the CMB very tightly constraints the difference $\dom$ (discussed further below). All the numerical constraints for the joint analysis are presented in Tab.~\ref{tab.results}; those for the key parameters are
\begin{equation}
  \begin{aligned}  
 (\omgeom, \wgeom) &=
  (0.3005_{-0.0081}^{+0.0069},
  -1.008_{-0.044}^{+0.047}) 
\\[0.2cm] 
  (\omgrow, \wgrow) &= 
  (0.3005_{-0.0073}^{+0.0071},
  -1.049_{-0.071}^{+0.086})
 \end{aligned}
 \quad \text{(split wCDM)}.
\end{equation}

Clearly, both pairs of parameters --- $\omgeom$ and $\omgrow$, as well as $\wgeom$ and $\wgrow$ --- are in good agreement with one another. Moreover, both equations of state are consistent (at 95\% confidence level) with the \lcdm\ expectation of $w=-1$.
Here, in parallel with our definition of $\dom$ in Eq.~(\ref{eq:dom}), we introduce the analogous quantity for $w$
\begin{equation}
\dw = \wgrow-\wgeom.
\label{eq:dw}
\end{equation}

\begin{figure*}[!t]
\centering
\includegraphics[width=0.8\textwidth]{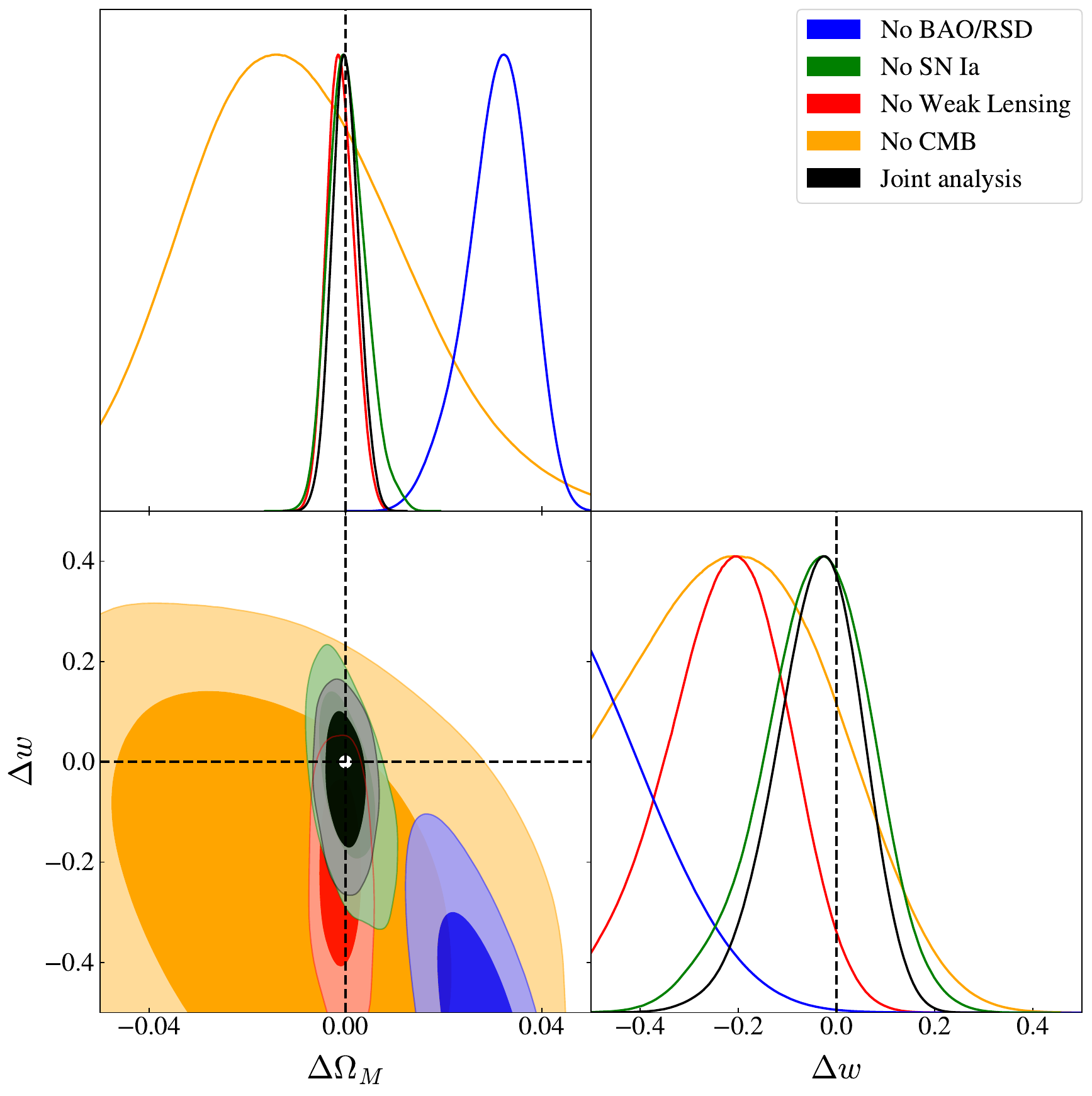}
\caption{Constraints on $\dom\equiv\omgrow-\omgeom$, $\dw\equiv \wgrow-\wgeom$ in the split wCDM model. We show the joint constraints, as well as cases when one probe at a time has been \textit{removed} from the analysis.}  
\label{fig:S8_dom_dw}
\end{figure*}

Figure \ref{fig:S8_dom_dw} shows the triangle plot for the parameters $\dom$ and $\dw$, marginalized over all other parameters. We show 68\% and 95\% contours from the joint analysis, as well as cases where one probe at the time has been \textit{removed} from the analysis. The combined-probes' constraints on the difference between the geometry and growth values of both $\Omega_M$ and $w$ are
\begin{equation}
  \begin{aligned}  
 \dom &= -0.0000_{-0.0027}^{+0.0026} 
\\[0.2cm] 
  \dw &= -0.041_{-0.084}^{+0.099}
 \end{aligned}
 \quad \text{(split wCDM)}.
\end{equation}
As in the split \LCDM\ model, we find no evidence for departures from the standard model, as we are consistent with $\dom=\dw=0$. We assess the significance of the pull away from zero by using Eq.~\eqref{pvalue} for $\dom$, and analogously for $\dw$. We find the respective significances of $1.0\sigma$ $(p=0.3072)$ for $\dom>0$, and $1.6\sigma$ $(p=0.1172)$ for $\dw>0$.

We would also like to investigate the individual constraints from the cosmological probes on the parameters in the split wCDM model, but because the individual probes provide weak constraints, we study the constraints when one probe at a time is \textit{removed} from the joint analysis. The results of these tests are shown in Table \ref{tab:noresults} and Fig.~\ref{fig:S8_dom_dw}.

%==================================================================

\begin{table*}[]
\centering
    \tiny
    \begin{tabular}{l|ccccccc} \hline\hline 
        Paper & $\omgeom$ & $\omgrow$ & $\dom$ & $\wgeom$ & $\wgrow$ & $\dw$\\
        \hline
        This work & $0.301 \pm 0.007$ & $0.302 \pm 0.006$ & $0.0004 \pm 0.0020$ & $-1.01 \pm 0.05$ & $-1.05 \pm 0.08$ & $-0.041 \pm 0.090$ \\
        \hline
        Wang et. al \citep{Wang:2007fsa} & --- & --- & $0.0044 \pm 0.0058$ & --- & --- & $-0.37 \pm 0.36$\\
        Ruiz \& Huterer \citep{Ruiz:2014hma} & $0.302 \pm 0.008$ & $0.321 \pm 0.017$ & $0.019^\star$ & $-1.13 \pm 0.06$ & $-0.77 \pm 0.08$ & $0.36^\star$\\
        Bernal et. al \citep{Bernal:2015zom} & $0.297 \pm 0.08$ & $0.29 \pm 0.08$ & $-0.007^\star$ & $-1.05 \pm 0.04$ & $-0.96 \pm 0.03$ & $0.09^\star$\\
        Muir et. al \citep{Muir:2020puy} & $0.304 \pm 0.008$ & $0.421 \pm 0.095$ & $-0.116 \pm 0.092$ & --- & --- & --- \\
        Ruiz-Zapatero et. al \citep{Ruiz-Zapatero:2021rzl} & $0.29\pm 0.02$ & $0.27\pm 0.04$ & $-0.02^\star$ & --- & --- & --- &\\
        \hline\hline 
    \end{tabular}
    
    \vspace{20pt}
    
\begin{tabular}{l|ccccc}
\hline\hline
\multirow{2}{*}{Paper}                                             & \multicolumn{5}{c}{Data}                                                                                                                                                                                                                                                                                                                                                                                                                 \\ \cline{2-6} 
                                                                   & CMB                                                                                                     & Weak lensing                                                                     & BAO/RSD                                                                                                   & SN Ia   & Galaxy clusters                                                                                                                    \\ \hline
This work                                                          & \begin{tabular}[c]{@{}c@{}}Planck 2018\\ (TTTEEE+lensing)\end{tabular}                                  & \begin{tabular}[c]{@{}c@{}}KiDS-1000\\ (COSEBIs)\end{tabular}                    & eBOSS DR16                                                                                                & Pantheon & ---                                                                                                                                \\ \hline
Wang et. al \cite{Wang:2007fsa}                  & \begin{tabular}[c]{@{}c@{}}WMAP3, ACBAR,\\ BOOMERanG and CBI\\ (TTTEEE)\end{tabular}                    & \begin{tabular}[c]{@{}c@{}}CTIO\\ (Aperture mass \\ statistic)\end{tabular}      & \begin{tabular}[c]{@{}c@{}}2dFGRS and \\ SDSS LRG\end{tabular}                                            & SNLS     & ---                                                                                                                                \\ \hline
Ruiz \& Huterer \cite{Ruiz:2014hma}              & \begin{tabular}[c]{@{}c@{}}Planck 2013\\ (Shift parameter and\\ early universe priors)\end{tabular}     & \begin{tabular}[c]{@{}c@{}}CFHTLens\\ (Shear 2PCF $\xi_{\pm}^{ij}$)\end{tabular} & \begin{tabular}[c]{@{}c@{}}6dFGS, SDSS LRG \\ and BOSS CMASS\end{tabular}                                 & SNLS     & \begin{tabular}[c]{@{}c@{}}MaxBCG\\  (Cluster counts)\end{tabular}                                                              \\ \hline
Bernal et. al \cite{Bernal:2015zom}              & \begin{tabular}[c]{@{}c@{}}Planck 2015\\ (TTTEEE+lensing)\end{tabular}                                  & ---                                                                              & \begin{tabular}[c]{@{}c@{}}6dFGS, SDSS-MGS,\\ BOSS-LOWZ, \\ BOSS-CMASS\\ and BOSS-Ly$\alpha$\end{tabular} & JLA      & \begin{tabular}[c]{@{}c@{}}Chandra X-ray \\ and Planck tSZ\\ $\left(\sigma_8\left(\frac{\Omega_M}{\alpha}\right)^{\beta}\right)$\end{tabular} \\ \hline
Muir et. al \cite{Muir:2020puy}                  & \begin{tabular}[c]{@{}c@{}}Planck 2015\\ (Shift parameter)\end{tabular}                                 & \begin{tabular}[c]{@{}c@{}}DES Y1\\ (Shear 2PCF $\xi_{\pm}^{ij}$)\end{tabular}   & \begin{tabular}[c]{@{}c@{}}DES Y1 and\\  BOSS DR12\end{tabular}                                           & DES Y1   & ---                                                                                                                                \\ \hline
Ruiz-Zapatero et. al \cite{Ruiz-Zapatero:2021rzl} & \begin{tabular}[c]{@{}c@{}}Planck 2018\\ (Shift parameter and\\ primordial power spectrum)\end{tabular} & \begin{tabular}[c]{@{}c@{}}KiDS-1000\\ (Band powers \\ spectrum) \end{tabular}               & \begin{tabular}[c]{@{}c@{}}6dFGS, BOSS DR12 \\ and BOSS DR14\end{tabular}                                 & ---      & ---                                                                                                                                \\ \hline\hline
\end{tabular}

\caption{\textit{Top:} Comparison of constraints from this work with existing literature. $\Omega_M$ constraints come from splitting only $\Omega_M$, and $w$ constraints from splitting both $w$ and $\Omega_M$. For works that do not quote $\dom$ and $\dw$, we estimate them from reported values of $\{\omgeom, \omgrow, \wgeom,\wgrow\}$. These estimates are denoted by $\star$, and their uncertainties are omitted as they cannot be computed from the provided results. Wang et. al \citep{Wang:2007fsa} only quoted $\dom$ and $\dw$, Muir et. al \citep{Muir:2020puy} could not constrain the $\wgeom - \wgrow$ plane, and Ruiz-Zapatero et. al \cite{Ruiz-Zapatero:2021rzl} did not study the \wcdm\ split, so all values are omitted. \textit{Bottom:} The various datasets for each probe used by all existing works. Missing values indicate a probe was not used in an analysis. Note that even if two works use the same dataset(s) for a probe, the variety in split implementation means that different parts of the data inform the growth and geometry constraints. For CMB, weak lensing and galaxy clusters, we identify in parenthesis the specific observable used in each work.}
    \label{tab:Comparisons}
\end{table*}

The most interesting result when removing one probe at a time is seen in the case when the BAO/RSD data is excluded from the joint analysis. In this case, the constraints are inconsistent with the standard model, and show $\dom=0.0305_{-0.0050}^{+0.0076}$ and $\dw=-0.88_{-0.25}^{+0.50}$. Converting the p-value into a Gaussian sigma, we find statistical significance at the level of $4.2 \sigma$ for $\dom<0$, and $3.6\sigma$ ($p=0.0004$) for $\dw>0$. 

This apparent discrepancy with the \lcdm\ expectation --- for constraints where the BAO/RSD dataset has been removed from the joint analysis --- can be understood as follows. In the ``No BAO/RSD'' case, the CMB and weak lensing largely inform the constraints on $\dom$ and $\dw$, as the SN Ia constrains only the geometry. Even in the unsplit \wcdm\ model, the CMB-only constraints do not perfectly agree with the standard model; they prefer lower values for $\Omega_M$ and a phantom ($w < -1$); see the middle and right panels of Fig.~\ref{fig:LCDM_wCDM}. This behavior is amplified in the split case, as can be seen in Figs.~\ref{fig:dom_dw} and~\ref{fig:omc-w-triangle}. Thus, it is not particularly surprising to see a discrepancy in the ``No BAO/RSD'' analysis.

Finally, regarding the joint analysis, the constraints on $\dom$ do not have any discernible correlation with $S_8$, while the departures of $\dw \neq 0$ occur only for high values of $S_8$ that are not allowed by the data; see Fig.~\ref{fig:dom_and_dw_and_S8} in the Appendix.

\renewcommand{\arraystretch}{1.5}
\begin{table*}[]
\centering
\begin{tabular}{l|c|c|c|c} \hline\hline
                      & $\Lambda$CDM                                    & $w$CDM                                            & split $\Lambda$CDM                & split $w$CDM       \\ \hline
$100~\omega_{b}$      & $2.246_{-0.013}^{+0.013}$                       & $2.246_{-0.014}^{+0.013}$                         & $2.249_{-0.015}^{+0.014}$         &  $2.250_{-0.015}^{+0.014}$         \\
$H_{0}$               & $68.87_{-0.36}^{+0.39}$                         & $68.90_{-0.79}^{+0.65}$                           & $68.58_{-0.65}^{+0.66}$              &  $68.67_{-0.85}^{+0.81}$           \\
$\ln10^{10}A_{s}$     & $3.040_{-0.014}^{+0.014}$                       & $3.040_{-0.014}^{+0.014}$                         & $3.042_{-0.015}^{+0.015}$         &  $3.039_{-0.015}^{+0.015}$         \\
$n_{s}$               & $0.9673_{-0.0037}^{+0.0033}$                    & $0.9672_{-0.0040}^{+0.0037}$                      & $0.9684_{-0.0042}^{+0.0039}$      &  $0.969_{-0.0042}^{+0.0045}$      \\
$\tau_{reio}$         & $0.0529_{-0.0070}^{+0.0068}$                    & $0.0526_{-0.0069}^{+0.0072}$                     & $0.054_{-0.0080}^{+0.0077}$     &  $0.0533_{-0.0073}^{+0.0079}$      \\
$\Omega_{M}^{\rm geom}$ & \multirow{2}{*}{$0.2998_{-0.0047}^{+0.0044}$}   & \multirow{2}{*}{$0.2997_{-0.0066}^{+0.0066}$}     & $0.3012_{-0.0071}^{+0.0066}$        &  $0.3005_{-0.0081}^{+0.0069}$      \\
$\Omega_{M}^{\rm grow}$ &                                                 &                                                   & $0.3017_{-0.0056}^{+0.0057}$         &    $0.3005_{-0.0073}^{+0.0071}$      \\
$w^{\rm geom}$          & \multirow{2}{*}{$-1$}                           & \multirow{2}{*}{$-1.002_{-0.024}^{+0.027}$}     & \multirow{2}{*}{$-1$}             &  $-1.008_{-0.044}^{+0.047}$        \\
$w^{\rm grow}$          &                                                 &                                                   &                                   &  $-1.049_{-0.071}^{+0.086}$          \\ \hline
$S_{8}$          & $0.8072_{-0.0088}^{+0.0081}$                    & $0.8074_{-0.0079}^{+0.0087}$     & $0.8081_{-0.0085}^{+0.0090}$      &  $0.8136_{-0.013}^{+0.013}$          \\ 
$\dom$    & $0$                                             & $0$               & $0.0004_{-0.0020}^{+0.0020}$                                                 & $-0.0000_{-0.0027}^{+0.0026}$  \\
$\Delta w$            & $0$                                             & $0$               & $0$                                                   & $-0.041_{-0.084}^{+0.099}$         \\ \hline\hline
\end{tabular}
\caption{The full parameter constraints (mean and $1\sigma$ uncertainties) for the standard \lcdm\, and \wcdm\ models as well as the split analogs to each. From top to bottom we show the following parameters: (i) the scaled baryon energy density $w_b = \Omega_b h^2$, (ii) the Hubble expansion rate at the present epoch, (iii) the amplitude of the primordial curvature power spectrum at $k_{\rm piv} = 0.05\, {\rm Mpc}^{-1}$, (iv) spectral index of density fluctuations, (v) optical depth to reionization, (vi - vii) the geometry and growth counterparts for the total matter energy density, (viii - ix) the geometry and growth counterparts for the dark energy equation of state. This is a fixed parameter $\wgeom = \wgrow = -1$ in the split \lcdm\ models. (x) the amplitude of density fluctuations on a scale $8 \hinvmpc$ then scaled by $\omgrow$: $S_8 = \sigma_8(\omgrow/0.3)^{0.5}$, (xi - xii) the differences between the geometry and growth parameters, $\dom \equiv \omgrow-\omgeom$ and $\dw \equiv \wgrow-\wgeom$.}
\label{tab.results}
\end{table*}

\section{Discussion}\label{sec:discussion}

There are a number of previous works on geometry-growth split analyses, each using different data (see Table~\ref{tab:Comparisons}) and split implementations (see Sec.~\ref{sec:split_def}). Such differences prevent a straightforward, unambiguous comparison across the literature, but there are still some trends worth highlighting.

When splitting only $\Omega_M$ (or $\Omega_{\Lambda}$, which is an equivalent split choice for a flat universe), almost all studies find no preference for $\dom \neq 0$. The one exception is Ref.~\citep{Bernal:2015zom}, who found a $\approx 4\sigma$ departure from $\dom = 0$, but they point out that they could not properly isolate the impact of systematics, particularly from the galaxy cluster and RSD probes, on this discrepancy. 

When splitting both $\Omega_M$ and $w$, however, the picture is more interesting. Two previous works have found $\wgrow > \wgeom$ \citep{Ruiz:2014hma,Bernal:2015zom}, meaning the data prefer less growth at late times than predicted by the standard model. Our joint analysis results, on the other hand, are consistent with $\wgrow = \wgeom$. This preference could arise from differences in data between the analyses; the growth constraints in previous works are informed primarily by probes whose theoretical modeling is challenging --- RSD data in \citep{Ruiz:2014hma}, and both galaxy cluster and RSD data in \citep{Bernal:2015zom} --- whereas in our work stronger constraints from the easier-to-model CMB, weak lensing, and BAO accentuates their contribution to the overall information content. Interestingly, our constraints from just BAO/RSD \textit{do} prefer $\wgrow > \wgeom$ (blue contour, right panel, Fig.~\ref{fig:dom_dw}), but this preference vanishes with the addition of other data. The CMB constraints in particular show contrary behavior to BAO/RSD, with a preference for $\wgrow < \wgeom$. This behavior is also consistent with Wang et.\ al \citep{Wang:2007fsa}, who used the same split implementation as us in CMB physics, but much older CMB data. Note, however, that these features in both the CMB and the BAO/RSD analyses are found at low significance ($< 2\sigma$). 

Omitting BAO/RSD from our joint analysis leads to constraints that display $4.2\sigma$  and $3.6\sigma$ deviations away from $\dom = 0$ and $\dw = 0$, respectively. These constraints are primarily informed by the CMB, which is known to show preference for phantom cosmological model in wCDM; for example, the Planck constraint on the equation of state in the \texttt{base\_w\_plikHM\_TTTEEE\_lowl\_lowE\_post\_lensing} model is $w=-1.57^{+0.16}_{-0.33}$. Adding BAO/RSD data, however,  both strengthens the constraints dramatically and returns them to agree well with \lcdm; see the third panel of Fig.~\ref{fig:LCDM_wCDM}. Ruiz \& Huterer \cite{Ruiz:2014hma} also found similar behaviors for datasets and split implementations that are different in detail from those of this work.

Finally, we address our decision to combine different datasets --- notably KiDS-1000 and Planck --- despite possible statistical inconsistencies between them. The nice feature of our geometry-growth test is that, if interpreted purely as a consistency test of the standard cosmological model, it is reasonably robust to assumptions about internal consistency of the data. This is because a failure to recover the standard model (here, $\omgeom=\omgrow$ and $\wgeom=\wgrow$) would indicate \textit{some} inconsistency in the underlying standard cosmological model (or else unaccounted-for systematics), regardless of assumptions about  internal  consistency of the probes. 

%==================================================================

\section{Summary and Conclusion}\label{sec:concl}

In this work, we consider a geometry-growth extension to the \lcdm\, and \wcdm\ cosmological standard models to check (i) if the different probes prefer models that deviate from GR, and (ii) whether such deviations can resolve the current $S_8 - \Omega_M$ tension between the CMB and large-scale structure. To constrain our extension model we use a diverse set of probes, including SN Ia, BAO, RSD, CMB, and weak lensing. While individual probes may often provide only weak constraints on the parameters of interest, combining the probes enables powerful degeneracy breaking and leads to precise constraints on the parameters that describe the geometry and the growth of structure in the universe.

Our model parameterization involves the split of
late-time cosmological parameters that describe dark energy into a pair of sets: one set captures the geometry from the expansion history while the other set describes the growth of structure. We consider split analyses of two cosmological models: flat \lcdm\, and flat wCDM model.  When we ``split'' the \lcdm\ model, we replace the matter density $\Omega_M$ with two parameters, ($\omgeom$, $\wgeom$), while for the wCDM model, we replace ($\Omega_M, w$) with four parameters: ($\omgeom, \omgrow, \wgeom, \wgrow$). We assign the geometry parameters to theoretical modeling of the distances and geometric projections (in lensing kernels), while we use the growth parameters to model the sound horizon at the drag epoch, the matter power spectrum, and the CMB source functions. While there exist multiple ways to implement the geometry-growth split in detail \citep{Wang:2007fsa, Ruiz:2014hma, Bernal:2015zom, Muir:2020puy},  all implementations are equally valid when performing a consistency check of GR: if GR holds, then the null-hypothesis, $\omgrow = \omgeom$ and $\wgrow = \wgeom$, must be satisfied regardless of the specific implementation of the split. 

We pay particular attention to constraints  on the parameter differences $\dom\equiv\omgrow-\omgeom$ and $\dw=\wgrow-\wgeom$ that describe departures from the standard model. In our implementation of the parameter split, we find that the parameter difference $\dom$ is particularly well constrained both from the CMB alone, and from all probes combined, indicating an excellent ability to constrain departures from $\Lambda$/wCDM.
 
In our split \lcdm\, analysis, the posteriors for the joint analysis are consistent with $\dom = 0$, and show no preference for departures from GR (Figure \ref{fig:om_split}).  The CMB provides the strongest constraint for this parameter, but all the other individual probes are also consistent with $\dom = 0$. In the $S_8 - \dom$ plane, we find that the CMB constraints are anti-correlated --- $S_8$ decreases with increasing $\dom$ --- while weak lensing constraints show no correlation. This implies that there exists a value of $\dom > 0$ for which the CMB constraints on $S_8$ would be brought into agreement with weak lensing. 
However, the weak lensing constraints in this parameter space are broad. Thus we find that the current datasets do not have enough constraining power to inform whether the geometry-growth split of the \lcdm\ standard model can resolve the $S_8$ - $\Omega_M$ tension between CMB and weak lensing. 

In the split \wcdm\ analysis, the joint-probe posteriors are entirely consistent with $\dom = 0$ and $\dw = 0$. Interestingly, the individual constraints from BAO/RSD prefer $\wgrow > \wgeom$ at $\approx 1.4\sigma$, which is a feature seen in previous works with different data and split implementations \citep{Ruiz:2014hma, Bernal:2015zom}. Removing BAO/RSD from the joint analysis results in an even more anomalous result, with $\dom > 0$ at $4.2 \sigma$ significance, and $\dw < 0$ at $3.6\sigma$. We conjecture that constraints from the combination of SN Ia, CMB and weak lensing are primarily informed by the CMB, and thus inherit the well-known preference of Planck's data for a phantom ($w<-1$) cosmological model in wCDM, which now manifests itself as a preference for nonzero $\dom$ and $\dw$. That preference goes away when BAO/RSD is added to the analysis.

Overall, we find that the standard cosmological model passes another test --- though just barely. 
It will be very interesting to see how the geometry-growth split constraints improve and evolve as new, better data are added.
Particularly interesting will be seeing how different cosmological probe evolve in relation to one another.
High quality data from upcoming Stage III (DES Y6, Hetdex, HSC) and Stage IV (J-PAS, DESI, LSST, WFIRST, Euclid, SKA) surveys will enable such exciting tests of the standard cosmological model.

\begin{acknowledgments}
It is a pleasure to thank Andreu Font Ribera and Andrei Cuceo for their help with the BAO QSO likelihoods, and then Marika Asgari, Tilman Troester and Benjamin Stölzner both for providing early access to \textsc{KCAP} (KiDS Cosmological Analysis Pipeline) and the corresponding Monte-Python likelihood, and for their consequent help with implementing these pipelines. UA acknowledges financial support from CAPES (Grants No.~PDSE-88881.361805/2019-01) and FAPERJ, and thanks the University of Michigan for hospitality.
DA is supported by the National Science Foundation Graduate Research Fellowship under Grant No. DGE 1746045. 
RvM acknowledges support from the Programa de Capacitação Institucional PCI/ON/MCTI.
DH has been supported by DOE under Contract No. DE-FG02-95ER40899,  NSF under contract AST-1812961, and NASA under contract 19-ATP19-0058. He thanks the Max Planck Institute for Astrophysics for hospitality. JSA acknowledges support from CNPq (Grants No.~310790/2014-0 and 400471/2014-0) and FAPERJ (Grant No.~204282).
Part of the computations were performed at the Virgo Cluster at Cosmo-ufes/UFES, which is funded by FAPES and administrated by Renan Alves de Oliveira, and at the National Observatory Data Center (DCON). This research was supported in part through computational resources and services provided by Advanced Research Computing (ARC), a division of Information and Technology Services (ITS) at the University of Michigan, Ann Arbor.
\end{acknowledgments}

\appendix

\section{More details about the constraints} 
\label{appx:details}

While our main results focus primarily on the split parameters, $\omgeom$, $\omgrow$, $\wgeom$, and $\wgrow$, we also constrain all the other parameters of the standard cosmological model. Here we present the constraints for all parameters (including those presented in our main analysis) for both the split \LCDM\ and split \wcdm\ models. All contours show $68\%$ and $95\%$ confidence intervals. For each 2D posterior, all other cosmological and nuisance parameters have been marginalized over.

\begin{figure*}[h!]
\centering
\includegraphics[width=0.67\textwidth]{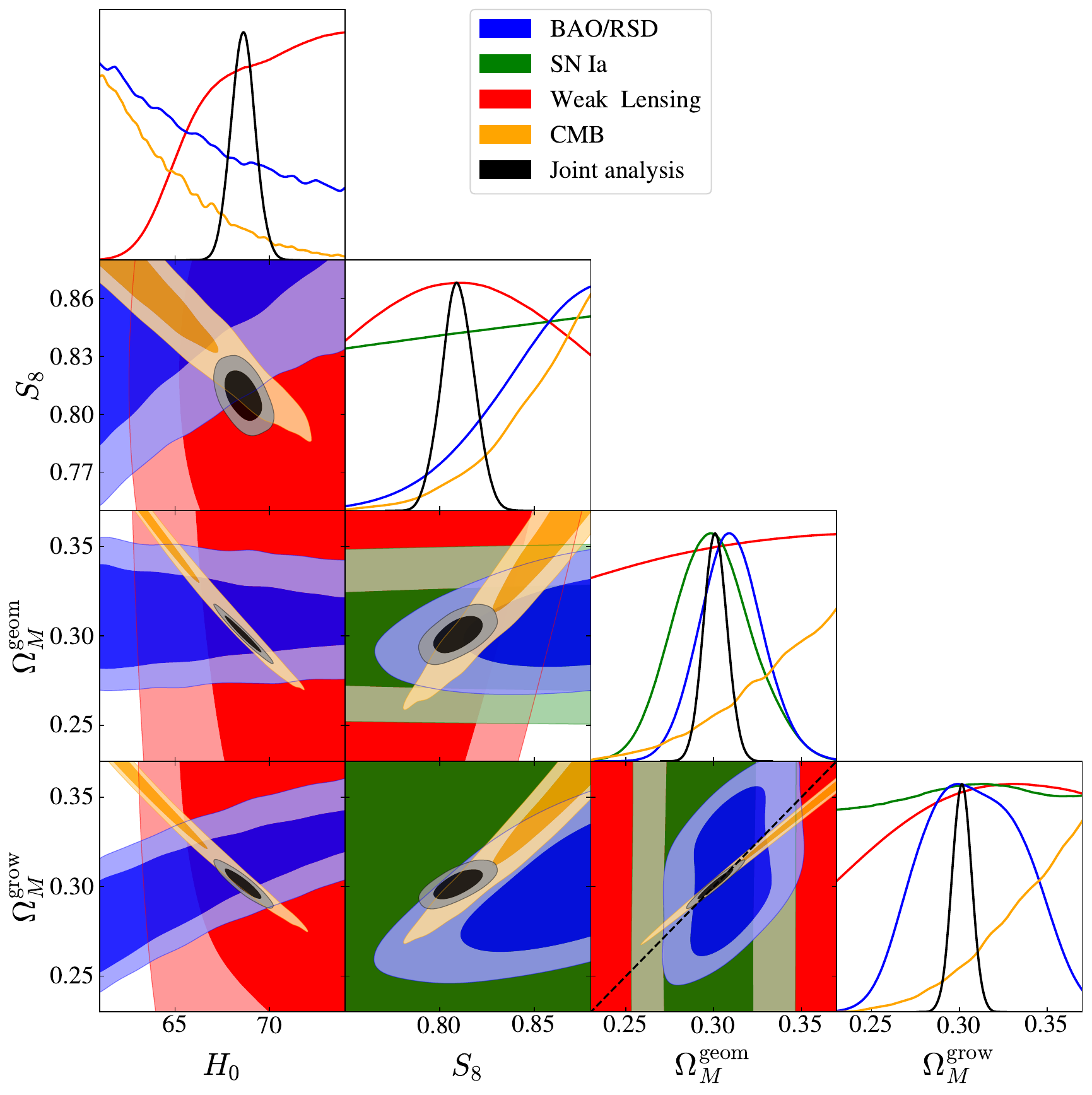}
\caption{Constraints in the split \lcdm\ model, showing the 68\% and 95\% limits on the parameters $H_{0}$, $S_{8}$, $\omgeom$, and $\omgrow$. The different-color contours show constraints from individual probes, and the black contours show the joint constraint.}
\label{fig:omctriangle}
\end{figure*}

Fig.~\ref{fig:omctriangle} shows the posteriors of $H_0$, $S_8$, $\omgeom$, and $\omgrow$. As expected, the geometry-growth split introduces new degeneracies leading to individual constraints that are considerably weaker compared to the fiducial $\Lambda$CDM model. For example, the CMB constraint on $H_0$ drastically weakens relative to that of the \LCDM\ model (and is also very non-Gaussian now), and the same happens to both the CMB and weak lensing constraints on $S_8$ as well. 

Nevertheless, there are some success stories even among the individual-constraint cases; for example, SN Ia individually constrain $\omgeom$ very well, while BAO/RSD constrains both $\omgeom$ and $\omgrow$.
Moreover, weak lensing, while unable to provide strong constraints on its own, helps break the degeneracy between $H_0$ and $\omgrow$ as measured by the CMB. Finally, the combination of cosmological probes in the full dimensional parameter space once again breaks degeneracies, and the joint constraints are precise for all parameters shown in  Fig.~\ref{fig:omctriangle}.

\begin{figure*}[h!]
\centering
\includegraphics[width=\textwidth]{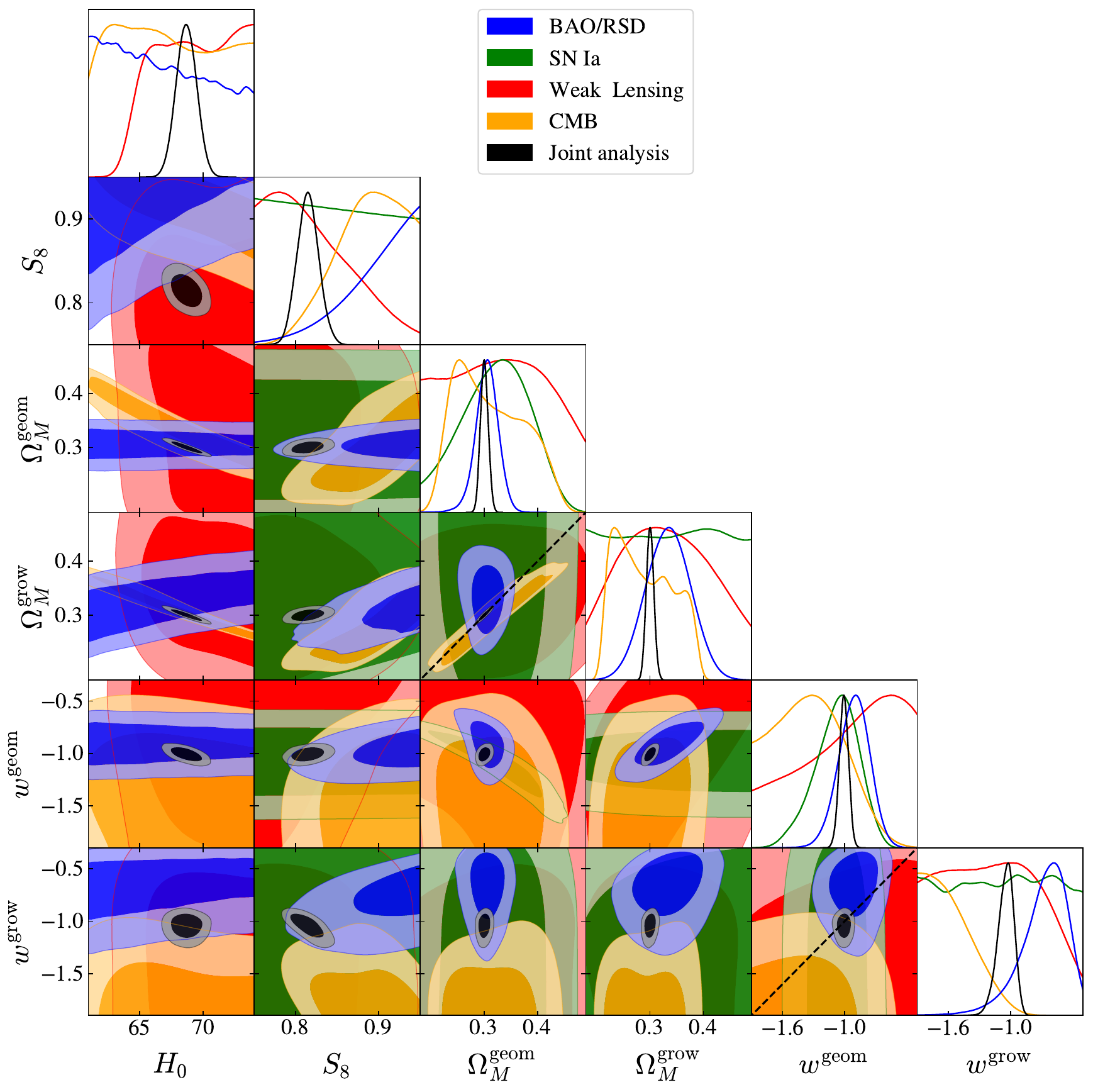}
\caption{Constraints in the split \wcdm\ model, showing the 68\% and 95\% limits on the parameters $H_{0}$, $S_{8}$, $\omgeom$, $\omgrow$, $\wgeom$, and $\wgrow$. The different-color contours show constraints from individual probes, and the black contours show the joint constraint. Note that the joint constraint in the $\omgeom$-$\omgrow$ plane is extremely thin and difficult to see (recall that we constrain $\dom$ really well), and largely overlaps with a few dashes of the diagonal line.
}
\label{fig:omc-w-triangle}
\end{figure*}

Fig.~\ref{fig:omc-w-triangle} shows the posteriors of $H_0$, $S_8$, $\omgeom$, $\omgrow$, $\wgeom$ and $\wgrow$. There are large degeneracies in the constraints of individual probes, indicating the intrinsic difficulty in constraining this parameter space \cite{Ruiz:2014hma}.
Nevertheless, we observe that the purely geometrical SN Ia data constrain the geometry parameters reasonably well, while BAO/RSD measurements give a respectable individual constraint on both geometry and growth parameters. 
When the probes are combined, we obtain a fairly precise constraint on most cosmological parameters.

\begin{figure*}[h!]
\centering
\includegraphics[width=0.49\textwidth]{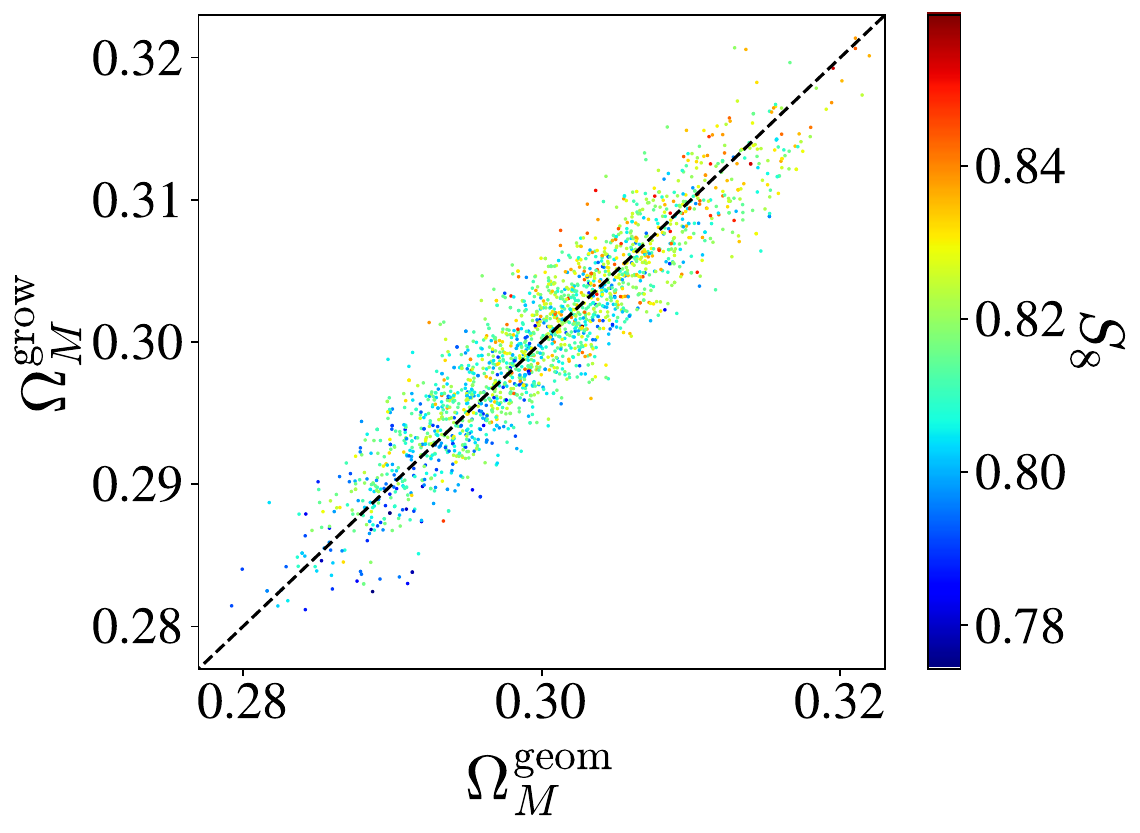}
\includegraphics[width=0.49\textwidth]{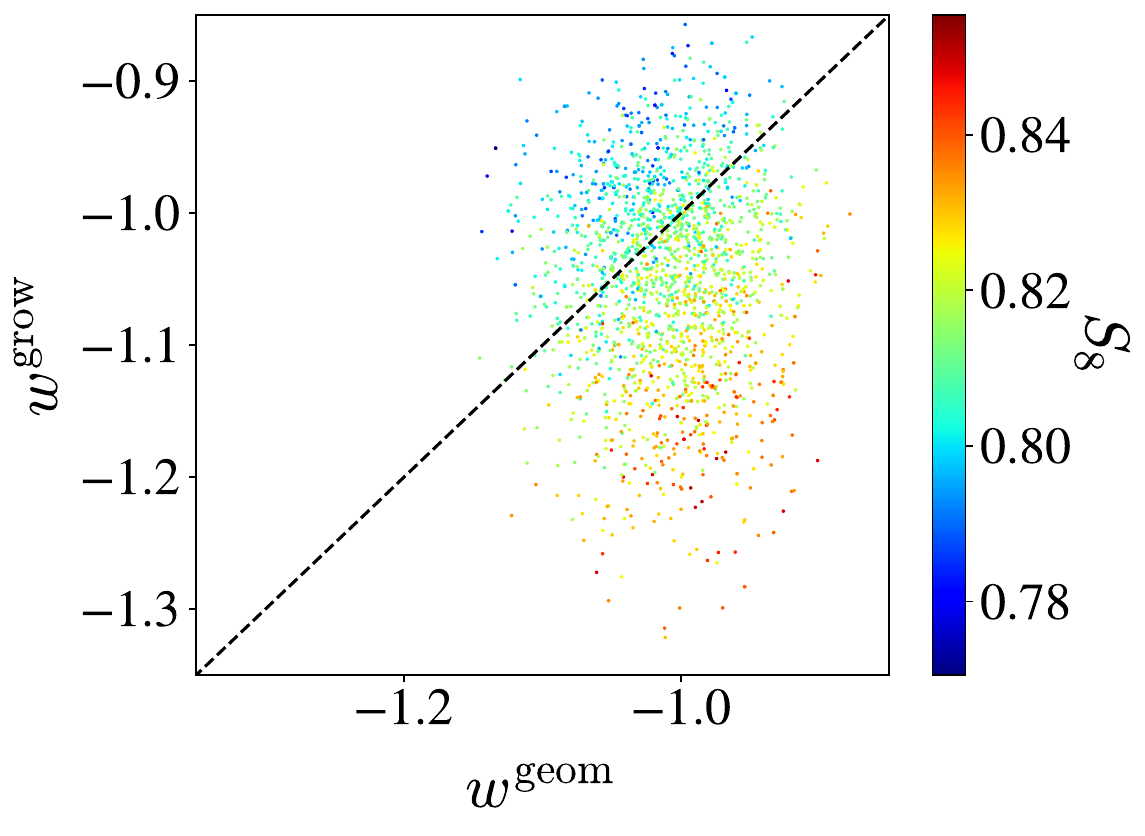}
\caption{Similar as the right panel of Fig.~\ref{fig:om_split1}, but for the split \wcdm\ model. \textit{Left panel:} Points from the posterior in the $\omgeom - \omgrow$ plane, color-coded by the value of $S_{8}$. \textit{Right panel:} Points from the posterior in the $\wgeom - \wgrow$ plane, color-coded by $S_{8}$. }
\label{fig:dom_and_dw_and_S8}
\end{figure*}

Finally, we comment on the tightness of the combined constraint in the split wCDM model. Fig.~\ref{fig:dom_and_dw_and_S8} shows a subsampling of Markov chains in the plane with $(\omgeom, \omgrow)$ (left panel) and $(\wgeom, \wgrow)$ (right panel). In each plane, we show representative values of the two split parameters, with colors showing the $S_8$ value for each point.  As in the split \LCDM\ model, we observe that the difference $\dom=\omgrow-\omgeom$ is very well constrained, without a noticeable dependence on the value of $S_8$. In contrast, $\wgeom$ is much better determined than $\wgrow$ and the two parameters are not particularly correlated. The difference $\dw\equiv\wgrow-\wgeom$ is nonzero only for cases when $S_8$ is considerably higher than its concordance value of $S_8\simeq 0.8$.

%=================================================================
%=================================================================

%=================================================================

%=================================================================
%=================================================================

\begin{table*}[]
\centering
\begin{tabular}{lcccc}
\hline\hline
\multicolumn{5}{c}{$\Lambda$CDM split}                                                                                                                                                                                                   \\ \hline\hline
\multicolumn{1}{l|}{}                        & \multicolumn{1}{c|}{No BAO/RSD}                     & \multicolumn{1}{c|}{No SN Ia}                     & \multicolumn{1}{c|}{No Weak Lensing}               & \multicolumn{1}{c}{No CMB} \\ \hline
\multicolumn{1}{l|}{$100~\omega_{b}$}        & \multicolumn{1}{c|}{$2.245_{-0.014}^{+0.015}$}     & \multicolumn{1}{c|}{$2.248_{-0.015}^{+0.015}$}     & \multicolumn{1}{c|}{$2.236_{-0.015}^{+0.013}$}     & $2.484_{-0.029}^{+0.140}$   \\
\multicolumn{1}{l|}{$H_{0}$}                 & \multicolumn{1}{c|}{$71.2_{-1.7}^{+1.5}$}         & \multicolumn{1}{c|}{$68.67_{-0.72}^{+0.67}$}       & \multicolumn{1}{c|}{$68.02_{-0.64}^{+0.66}$}       & $68.03_{-1.30}^{+0.47}$     \\
\multicolumn{1}{l|}{$\ln10^{10}A_{s}$}      & \multicolumn{1}{c|}{$3.044_{-0.015}^{+0.014}$}     & \multicolumn{1}{c|}{$3.041_{-0.013}^{+0.014}$}     & \multicolumn{1}{c|}{$3.052_{-0.015}^{+0.014}$}     & $3.27_{-0.16}^{+0.22}$    \\
\multicolumn{1}{l|}{$n_{s}$}                 & \multicolumn{1}{c|}{$0.9674_{-0.0042}^{+0.0043}$}  & \multicolumn{1}{c|}{$0.9678_{-0.0045}^{+0.0039}$}  & \multicolumn{1}{c|}{$0.9651_{-0.0036}^{+0.0048}$}  & $0.989_{-0.096}^{+0.099}$ \\
\multicolumn{1}{l|}{$\tau_{reio}$}           & \multicolumn{1}{c|}{$0.0546_{-0.0078}^{+0.0070}$}  & \multicolumn{1}{c|}{$0.0536_{-0.0073}^{+0.0074}$} & \multicolumn{1}{c|}{$0.0571_{-0.0092}^{+0.0065}$}  & ---                          \\
\multicolumn{1}{l|}{$\Omega_{M}^{\rm geom}$} & \multicolumn{1}{c|}{$0.275_{-0.015}^{+0.014}$}    & \multicolumn{1}{c|}{$0.3004_{-0.0076}^{+0.0069}$}  & \multicolumn{1}{c|}{$0.3090_{-0.0079}^{+0.0066}$}   & $0.295_{-0.014}^{+0.013}$ \\
\multicolumn{1}{l|}{$\Omega_{M}^{\rm grow}$} & \multicolumn{1}{c|}{$0.281_{-0.012}^{+0.012}$}     & \multicolumn{1}{c|}{$0.3011_{-0.0060}^{+0.0056}$}   & \multicolumn{1}{c|}{$0.3098_{-0.0064}^{+0.0056}$}  & $0.279_{-0.015}^{+0.012}$ \\ \hline
\multicolumn{1}{l|}{$S_{8}$}                 & \multicolumn{1}{c|}{$0.789_{-0.013}^{+0.013}$}    & \multicolumn{1}{c|}{$0.8077_{-0.0087}^{+0.0090}$}   & \multicolumn{1}{c|}{$0.828_{-0.010}^{+0.010}$}    & $0.780_{-0.017}^{+0.017}$ \\
\multicolumn{1}{l|}{$\dom$}                  & \multicolumn{1}{c|}{$0.0057_{-0.0031}^{+0.0035}$}& \multicolumn{1}{c|}{$0.0008_{-0.0021}^{+0.0022}$}& \multicolumn{1}{c|}{$0.0008_{-0.0020}^{+0.0020}$} & $-0.016_{-0.016}^{+0.012}$ \\ \hline\hline
\multicolumn{5}{c}{wCDM split}                                                                                                                                                                                                           \\ \hline\hline
\multicolumn{1}{l|}{}                        & \multicolumn{1}{c|}{No BAO/RSD}                     & \multicolumn{1}{c|}{No SN Ia}                     & \multicolumn{1}{c|}{No Weak Lensing}               & No CMB                     \\ \hline
\multicolumn{1}{l|}{$100~\omega_{b}$}        & \multicolumn{1}{c|}{$2.250_{-0.015}^{+0.016}$}      & \multicolumn{1}{c|}{$2.251_{-0.016}^{+0.014}$}     & \multicolumn{1}{c|}{$2.244_{-0.014}^{+0.016}$}     & $2.456_{-0.038}^{+0.170}$   \\
\multicolumn{1}{l|}{$H_{0}$}                 & \multicolumn{1}{c|}{$78.1_{-1.4}^{+3.5}$}         & \multicolumn{1}{c|}{$68.6_{-1.6}^{+1.4}$}         & \multicolumn{1}{c|}{$67.96_{-0.74}^{+0.77}$}       & $68.35_{-1.70}^{+0.60}$      \\
\multicolumn{1}{l|}{$\ln10^{10}A_{s}$}       & \multicolumn{1}{c|}{$3.036_{-0.015}^{+0.015}$}     & \multicolumn{1}{c|}{$3.038_{-0.015}^{+0.015}$}     & \multicolumn{1}{c|}{$3.043_{-0.015}^{+0.014}$}     & $3.03_{-0.24}^{+0.32}$    \\
\multicolumn{1}{l|}{$n_{s}$}                 & \multicolumn{1}{c|}{$0.9687_{-0.0043}^{+0.0042}$}  & \multicolumn{1}{c|}{$0.9688_{-0.0045}^{+0.0042}$}  & \multicolumn{1}{c|}{$0.9670_{-0.0043}^{+0.0045}$}   & $1.01_{-0.10}^{+0.10}$      \\
\multicolumn{1}{l|}{$\tau_{reio}$}           & \multicolumn{1}{c|}{$0.0520_{-0.0077}^{+0.0076}$} & \multicolumn{1}{c|}{$0.0528_{-0.0076}^{+0.0077}$} & \multicolumn{1}{c|}{$0.0541_{-0.0083}^{+0.0072}$} & ---                          \\
\multicolumn{1}{l|}{$\Omega_{M}^{\rm geom}$} & \multicolumn{1}{c|}{$0.203_{-0.026}^{+0.012}$}    & \multicolumn{1}{c|}{$0.301_{-0.011}^{+0.011}$}    & \multicolumn{1}{c|}{$0.301_{-0.011}^{+0.011}$}    & $0.297_{-0.018}^{+0.017}$ \\
\multicolumn{1}{l|}{$\Omega_{M}^{\rm grow}$} & \multicolumn{1}{c|}{$0.2330_{-0.0200}^{+0.0085}$}     & \multicolumn{1}{c|}{$0.301_{-0.013}^{+0.013}$}    & \multicolumn{1}{c|}{$0.3086_{-0.0070}^{+0.0072}$}   & $0.287_{-0.019}^{+0.014}$ \\
\multicolumn{1}{l|}{$w^{\rm geom}$}          & \multicolumn{1}{c|}{$-0.804_{-0.041}^{+0.058}$}   & \multicolumn{1}{c|}{$-1.004_{-0.085}^{+0.096}$}    & \multicolumn{1}{c|}{$-1.014_{-0.045}^{+0.043}$}    & $-1.004_{-0.066}^{+0.073}$ \\
\multicolumn{1}{l|}{$w^{\rm grow}$}          & \multicolumn{1}{c|}{$-1.69_{-0.22}^{+0.48}$}      & \multicolumn{1}{c|}{$-1.043_{-0.069}^{+0.087}$}    & \multicolumn{1}{c|}{$-1.25_{-0.11}^{+0.14}$}      & $-1.36_{-0.18}^{+0.40}$    \\ \hline
\multicolumn{1}{l|}{$S_{8}$}                 & \multicolumn{1}{c|}{$0.934_{-0.048}^{+0.045}$}     & \multicolumn{1}{c|}{$0.812_{-0.017}^{+0.016}$}    & \multicolumn{1}{c|}{$0.858_{-0.018}^{+0.018}$}    & $0.800_{-0.028}^{+0.032}$ \\
\multicolumn{1}{l|}{$\dom$}                  & \multicolumn{1}{c|}{$0.0305_{-0.0050}^{+0.0076}$}  & \multicolumn{1}{c|}{$0.0004_{-0.0044}^{+0.0032}$}& \multicolumn{1}{c|}{$-0.0011_{-0.0030}^{+0.0026}$}& $-0.016_{-0.024}^{+0.020}$ \\
\multicolumn{1}{l|}{$\Delta w$}              & \multicolumn{1}{c|}{$-0.88_{-0.25}^{+0.50}$}      & \multicolumn{1}{c|}{$-0.039_{-0.099}^{+0.120}$}   & \multicolumn{1}{c|}{$-0.23_{-0.12}^{+0.14}$}& $-0.35_{-0.19}^{+0.40}$ \\ \hline\hline
\end{tabular}
\caption{The full parameter constraints (mean and $1\sigma$ uncertainties) for the joint analyses after removing one particular probe. We show the same parameters as Tab. \ref{tab.results}.}
\label{tab:noresults}
\end{table*}


\end{document}